\documentclass{mn2ex} % preprint/draft versions
\usepackage{psfig}

\title[6dFGS: Sample and Redshifts] 
{The 6dF Galaxy Survey: Samples, Observational Techniques and the
First Data Release}

\author[Jones et al.]{
\parbox[t]{\textwidth}{
D.\ Heath Jones$^{1}$, Will Saunders$^{2}$, Matthew Colless$^{2}$,
Mike A.\ Read$^{3}$, \\
Quentin A.\ Parker$^{4,2}$, Fred G.\ Watson$^{2}$, Lachlan A.\ 
Campbell$^{1}$, Daniel Burkey$^{3}$, \\
Thomas Mauch$^{5}$, Malcolm Hartley$^{2}$, Paul Cass$^{2}$, Dionne
James$^{2}$,  Ken Russell$^{2}$, \\
Kristin Fiegert$^{2}$, John Dawe$^{2}$, John Huchra$^{6}$, Tom
Jarrett$^{7}$, Ofer Lahav$^{8}$, \\ 
John Lucey$^{9}$, Gary A.\ Mamon$^{10,11}$, Dominique Proust$^{11}$,
Elaine M.\ Sadler$^{5}$ \\ 
and Ken-ichi Wakamatsu$^{12}$ \\
}
\vspace*{6pt} \\
$^1$Research School of Astronomy \&; Astrophysics, The Australian
National University, \\ Weston Creek, ACT 2611, Australia
({\tt heath, lachlan@mso.anu.edu.au}) \\
$^2$Anglo-Australian Observatory, P.O.\ Box 296, Epping, NSW 2121,
Australia ({\tt will, colless, fgw@aao.gov.au})\\
$^3$Institute for Astronomy, Royal Observatory, Blackford Hill,
Edinburgh, EH9~3HJ, United Kingdom \\ 
$^4$Department of Physics, Macquarie University, Sydney 2109, Australia \\
$^5$School of Physics, University of Sydney, NSW 2006, Australia \\
$^6$Harvard-Smithsonian Center for Astrophysics, 60 Garden St  MS20,
Cambridge, MA 02138-1516, USA \\ 
$^7$Infrared Processing and Analysis Center, California Institute of
Technology, Mail Code 100-22, \\ 
770 South Wilson Avenue, Pasadena, CA 91125, USA\\
$^8$Institute of Astronomy, University of Cambridge, Madingley Road,
Cambridge CB3~0HA, United Kingdom \\ 
$^9$Department of Physics, University of Durham, South Road,
Durham DH1~3LE, United Kingdom \\
$^{10}$Institut d'Astrophysique de Paris (CNRS UMR 7095),
98 bis Bd Arago, F-75014 Paris, France\\
$^{11}$GEPI (CNRS UMR 8111), Observatoire de Paris, F-92195 Meudon, France\\
$^{12}$Faculty of Engineering, Gifu University, Gifu 501--1192, Japan\\
} \date{Accepted ---. Received ---; in original form ---.}

\newlength{\plotwidth}
\newlength{\fullwidth}
\newcommand{\eg}{\mbox{\it e.g.}}

\newcommand{\etal}{\mbox{et~al.}}
\newcommand{\bj}{\mbox{$b_{\rm\scriptscriptstyle J}$}}
\newcommand{\rf}{\mbox{$r_{\rm\scriptscriptstyle F}$}}
\newcommand{\perpix}{\mbox{\,pixel$^{-1}$}}

\newcommand{\invMpc}{\mbox{$\,h\,{\rm Mpc}^{-1}$}}

\newcommand{\kms}{\mbox{\,km\,s$^{-1}$}}

\newcommand{\zcat}{19\,570}
\newcommand{\twodf}{8\,444}
\newcommand{\lit}{28\,014}

\newcommand{\totalDB}{52\,048}     
\newcommand{\uniqueDB}{46\,474}     
\newcommand{\totalDBqual}{43\,945}           % i.e. Q>=3  
\newcommand{\uniqueDBqual}{39\,649}           % i.e. Q>=3
\newcommand{\aimsixdf}{150\,000}
\newcommand{\targets}{174\,442}

\newcommand{\jonesLF}{75\,000}

\newcommand{\fields}{524}
\newcommand{\totaltiles}{1564}
\newcommand{\tilesA}{547}
\newcommand{\tilesB}{595}
\newcommand{\tilesC}{422}
\newcommand{\plotone}[1]
    {\centering \leavevmode \psfig{file=#1,width=\plotwidth,clip=}}

\newcommand{\plotfull}[2]
    {\centering \leavevmode \psfig{file=#1,width=#2\fullwidth,clip=}}
\newcommand{\plotrot}[3]
    {\centering \leavevmode \psfig{file=#1,width=#2\fullwidth,angle=#3,clip=}}

\begin{document}
%\sloppy                % referee version

\maketitle

\begin{abstract}
  
  The 6dF Galaxy Survey (6dFGS) aims to measure the redshifts of around
  \aimsixdf\ galaxies, and the peculiar velocities of a 15\,000-member
  sub-sample, over almost the entire southern sky. When complete, it
  will be the largest redshift survey of the nearby universe, reaching
  out to about $z\sim0.15$, and more than an order of magnitude larger
  than any peculiar velocity survey to date.  The targets are all
  galaxies brighter than $K_{\rm tot} = 12.75$ in the 2MASS Extended
  Source Catalog (XSC), supplemented by 2MASS and SuperCOSMOS galaxies
  that complete the sample to limits of $(H, J, r_F, b_J) = (13.05,
  13.75, 15.6, 16.75)$. Central to the survey is the Six-Degree Field
  (6dF) multi-fibre spectrograph, an instrument able to record 150
  simultaneous spectra over the $5.7^\circ$-field of the UK Schmidt
  Telescope. An adaptive tiling algorithm has been employed to ensure
  around 95\% fibering completeness over the 17046~deg$^2$ of the southern sky
  with $|\,b\,|>10^\circ$. Spectra are obtained in two observations using
  separate V and R gratings, that together give $R \sim 1000$ over
  at least 4000 -- 7500\,\AA\ and signal-to-noise ratio $\sim 10$ per pixel. 
  Redshift measurements are obtained semi-automatically, and are assigned a
  quality value based on visual inspection. The 6dFGS database is
  available at {\tt http://www-wfau.roe.ac.uk/6dFGS/}, with public data
  releases occuring after the completion of each third of the survey.

\end{abstract}


\newpage

% ssssssssssssssssssssssssssssssssssssssssssssssssssssssssssssssssssssss

\section{INTRODUCTION}
\label{sec:introduction}

Wide-scale redshift surveys such as the 2dF Galaxy Redshift Survey and
Sloan Digital Sky Survey (2dFGRS, Colless \etal\ 2001b; SDSS, York \etal\ 
2001) have made significant advances in our understanding of the matter
and structure of the wider universe.  These include the precise
determination of the luminosity function of galaxies (\eg\ Folkes \etal\ 
1999, Cross \etal\ 2001, Cole \etal\ 2001, Blanton \etal\ 2001, Madgwick
\etal\ 2002, Norberg \etal\ 2002, Blanton \etal\ 2003), the space
density of nearby rich galaxy clusters (De Propris \etal\ 2002, Goto
\etal\ 2003), and large-scale structure formation and mass density
(Peacock \etal\ 2001, Percival \etal\ 2001, Efstathiou \etal\ 2002,
Verde \etal\ 2002, Lahav \etal\ 2002, Zehavi \etal\ 2002, Hawkins \etal\ 
2003, Szalay \etal\ 2003). 

While such surveys have also greatly refined our view of the local
universe, better determination of several key parameters can be made
where a knowledge of galaxy mass can be combined with redshift. The
2dFGRS and SDSS surveys are both optically-selected, inevitably biasing
them in favour of currently star-forming galaxies. The 2dF and Sloan
spectrographs also have fields of view too small to allow full sky
coverage in realistic timescales, limiting their utility for dynamical
and cosmographic studies. The 6dF Galaxy Survey (6dFGS) is a dual
redshift/peculiar velocity survey that endeavours to overcome the
limitations of the 2dFGRS and SDSS surveys in these areas.

The primary sample for the 6dFGS is selected in the $K_s$ band from
the 2MASS survey (Jarrett \etal\ 2000). The magnitudes used in the
selection are estimated total magnitudes. These features combined mean
that the primary sample is as unbiased a picture of the universe, in
terms of the old stellar content of galaxies, as is possible at the
current time. The near-infrared selection also enables the survey to
probe closer to the Galactic equator before extinction becomes an
issue; the survey covers the entire southern sky with $|\,b\,|>10^\circ$.

The redshift of a galaxy includes both recessional and peculiar velocity
components, so that a redshift survey alone does not furnish a true
three-dimensional distribution for the galaxies. However, by measuring
these components separately, it is possible to determine the
three-dimensional distributions of both the galaxies and the underlying
mass.

Observationally, significantly greater effort is required to obtain the
distances and peculiar velocities of galaxies than their redshifts.
Distance estimators based on the Fundamental Plane of early-type
galaxies (FP; Dressler \etal\ 1987, Djorgovski \& Davis 1987), as used
in the 6dFGS, require measurements of the galaxies' internal velocity
dispersions. Velocity dispersions demand a signal-to-noise ratio (S/N)
in the spectrum at least 3 to 5 times higher than redshift measurements.
Furthermore, the FP distance estimators need such spectroscopy to be
supported by photometry which can be used to determine the galaxies'
surface brightness profiles. 

We measure peculiar velocities as the discrepancy between the redshift
and the estimated distance. For realistic cosmologies, peculiar
velocities increase weakly with distance, but remain $<1000 \kms$. 
Redshift errors are small, and have little or no dependence on
distance. On the other hand, all existing distance estimators have
significant, intrinsic and fractional, uncertainties in distance; for
the FP estimators this uncertainty is typically about 20\% for a
single galaxy measurement. This linear increase in the errors with
distance, compared with more or less fixed peculiar velocities, means
that the uncertainty on a single peculiar velocity becomes dominated
by the intrinsic uncertainties at redshifts $cz \sim 5000
\kms$. Consequently, all previous peculiar velocity surveys have
traced the velocity field, and hence the mass distribution, only out
to distances of about 5000\kms\ (for relatively dense field samples of
individual galaxies; \eg\ Dressler \etal\ 1987, Giovanelli \etal\
1998, da Costa \etal\ 2000) or 15000\kms\ (for relatively sparse
cluster samples, where distances for multiple galaxies are combined;
\eg\ Lauer and Postman 1994, Hudson \etal\ 1999, Colless \etal\
2001a). The smaller volumes are highly subject to cosmic variance,
while the larger volumes are too sparsely sampled to reveal much
information about the velocity field. In order to differentiate
cosmological models, and constrain their parameters, both the survey
volume and galaxy sampling need to be significantly
increased. Moreover, since peculiar velocities amount to at most a few
per cent of the velocity over much of the volume, both the underlying
sample and the direct distance estimates have to be extraordinarily
homogeneous, preferably involving unchanging telescopes, instrumental
set ups and procedures in each case.

The two facets of the 6dF Galaxy Survey are a redshift survey of
\aimsixdf\ galaxies over the southern sky, and a
peculiar velocity survey of a subset of some 15\,000 of these galaxies.

The aims of the redshift survey are: (1)~To take a near-infrared
selected sample of galaxies and determine their luminosity function as a
function of environment and galaxy type.  From these, the stellar mass
function and mean stellar fraction of the local universe can be
ascertained and compared to the integrated stellar mass inferred from
measures of cosmic star formation history.
(2)~To measure galaxy
clustering on both small and large scales and its relation to stellar
mass. In doing so, the 6dFGS will provide new insight into the
scale-dependence of galaxy biasing and its relationship to dark matter.
(3)~To determine the power spectrum of this galaxy clustering on scales
similar to those spanned by the 2dF Galaxy Redshift Survey and Sloan
Digital Sky Surveys. (4)~To delineate the distribution of galaxies in
the nearby universe. The 6dFGS will not cover as many galaxies as either
the 2dF Galaxy Redshift Survey or Sloan Digital Sky Survey. However,
because it is a large-volume survey of nearby galaxies, it provides the
ideal sample on which to base a peculiar velocity survey. In this
respect, the complete 6dFGS will be more than ten times larger in number
and span twice the volume of the PSCz Survey (Saunders \etal\ 2000),
previously the largest survey of the local universe. (5)~To furnish a
complete redshift catalogue for future studies in this regime. As
discussed above, precise distance measures impose stronger demands on
signal-to-noise than do redshifts.  With such higher quality spectra, it
is also possible to infer properties of the underlying stellar
population, such as ages and chemical abundances. Having such
measurements along with the galaxy mass and knowledge of its local
envrionment, will afford unprecedented opportunities to understand the
processes driving galaxy formation and evolution.

This redshift catalogue will provide the basis for a volume-limited
sample of early-type galaxies for the peculiar velocity survey. The aims
of the peculiar velocity survey are: (1)~A detailed mapping of the
density and peculiar velocity fields to around 15000\kms\, over half the
local volume. (2)~The inference of the ages, metallicities and
star-formation histories of E/S0 galaxies from the most massive systems
down to dwarf galaxies. The influence of local galaxy density on these
parameters will also be of key interest to models of galaxy formation.
(3)~To understand the bias of galaxies (number density versus total mass
density field) and its variation with galaxy parameters and environment.

One novel feature of the 6dF Galaxy Survey compared to earlier redshift
and peculiar velocity surveys is its near-infrared source selection. The
main target catalogues are selected from the Two Micron All Sky Survey
(2MASS; Jarrett \etal\ 2000) using total galaxy magnitudes in $JHK$.
There are several advantages of choosing galaxies in these bands. First,
the near-infrared spectral energy distributions (SEDs) of galaxies are
dominated by the light of their oldest stellar populations, and hence,
the bulk of their stellar mass. Traditionally, surveys have selected
target galaxies in the optical where galaxy SEDs are dominated by
younger, bluer stars. Second, the E/S0 galaxies that will provide the
best targets for Fundamental Plane peculiar velocity measures represent
the largest fraction by galaxy type of near-infrared-selected samples.
Finally, the effects of dust extinction are minimal at long wavelengths.
In the target galaxies, this means that the total near-infrared
luminosity is not dependent on galaxy orientation and so provides a
reliable measure of galaxy mass. In our own Galaxy, it means the 6dFGS
can map the local universe nearer to the plane of the Milky Way than
would otherwise be possible through optical selection.

In this paper we describe the key components contributing to the
realisation of the 6dF Galaxy Survey. Section~\ref{sec:instrument}
describes the Six-Degree Field instrument; section~\ref{sec:design}
details the compilation of the input catalogues and the optimal
placement of fields and fibres on sources therein;
section~\ref{sec:implementation} outlines the methods used to obtain and
reduce the spectra, and derive redshifts from them; 
section~\ref{sec:datarelease} summarises the First Data Release 
of \uniqueDB\ unique galaxy redshifts and describes the 6dF Galaxy Survey
on-line database; section~\ref{sec:conclusions} provides concluding
remarks.

% ssssssssssssssssssssssssssssssssssssssssssssssssssssssssssssssssssssss

\section{THE SIX-DEGREE FIELD SPECTROGRAPH}
\label{sec:instrument}

Central to the 6dFGS is the Six-Degree Field multi-object fibre
spectroscopy facility, (hereafter referred to as 6dF), constructed by 
the Anglo-Australian Observatory (AAO) and operated on the United Kingdom 
Schmidt Telescope (UKST). This instrument
has three major components: (1) an $r$-$\theta$ robotic fibre positioner,
(2) two interchangeable 6dF field plates that contain the fibres to be
positioned, and (3) a fast Schmidt spectrograph that accepts the fibre-slit
from a 6dF field plate. The process of 6dF operation involves
configuring a 6dF field plate on target objects, mounting this
configured plate at the focal surface of the UKST and feeding the output
target fibre bundle into an off telescope spectrograph. While one target
field is being observed the other field plate can be configured.

6dF can obtain up to 150 simultaneous spectra across the
5.7\degr-diameter field of the UKST. Fuller descriptions of the 6dF
instrument have been given elsewhere, (Parker \etal\ 1998, Watson \etal\ 
2000, Saunders \etal\ 2001); here we summarise only those features
important to the Galaxy Survey.

The 6dF positioner, though building on the expertise and technology
successfully developed and employed in the AAO's 2dF facility (\eg\
Lewis \etal\ 2002), and in previous incarnations of fibre-fed
spectrographs at the UKST such as FLAIR (Parker \& Watson 1995), is
nevertheless a significant departure in both concept and
operation. 6dF employs an $r$-$\theta$ robotic fibre positioner
constructed as a working proto-type for the OzPoz fibre positioner
built under contract by the AAO and now commissioned on the European
Southern Observatory Very Large Telescope.
This fibre-placement technology can place fibre-buttons directly and
accurately onto the convex focal surface of the UKST, via a curved radial
arm matched to the focal surface. This is coupled with a complete
$>360^\circ$ $\theta$-travel and with a pneumatically controlled
fibre-gripper travelling in the $z$-direction.  Gripper positioning is
honed (to $<10\mu$m) using an inbuilt small CCD camera to permit
centroid measurement from back-illuminated images of each fibre.

Unlike 2dF, the 6dF positioning robot is off-telescope, in a
special enclosure on the dome floor. Two identical field plate units
are available, which allows one to be mounted on the telescope taking
observations whilst the other is being configured by the 6dF
robot. Each field plate contains a ring of 154 fibre buttons
comprising 150 science fibres and four bundles of guide fibres all
arranged around the curved field plate. Each 100$\mu$m (6.7\arcsec)
science fibre is terminated at the input end by a 5\,mm diameter
circular button, containing an elongated SF2 prism to deflect the
light into the fibre and a strong rare-earth magnet for adhesion to
the field plate. Targets closer than $5.7'$ on the sky cannot be 
simultaneously configured due to the clearances required to avoid 
collisions and interference between buttons. The buttons and trailing fibres
are incorporated into individual retractors housed within the main
body of the field plate under slight elastic tension. The 150 target
fibres feed into a fibre cable wrap 11 metres long, and terminate in a
fibre slit-block mounted in the 6dF spectrograph.

A full field configuration takes around 30-40 minutes depending on
target disposition and target numbers, plus about half this time to
park any prior configuration. This is less than the time spent by a
configured field plate in the telescope of 1.5 - 2.5 hours (depending
on conditions). There is however a $25-30$ minute overhead between
fields, needed for parking the telescope, taking arcs and flats,
manually unloading and loading the field plates, taking new arcs and
flats, and acquiring the new field. The acquisition is via four guide
fibres, each consisting of 7 $\times$ 100$\mu$m fibres hexagonally
packed to permit direct imaging from four guide stars across the field
plate. The guide fibres proved extremely fragile in use, and also hard
to repair until a partial redesign in 2002; as a result of this a
significant fraction of the data was acquired with three, or
occasionally even two, guide fibres, with consequent loss of
acquisition accuracy and signal-to-noise.

The spectrograph is essentially the previous bench-mounted FLAIR II
instrument\footnote{Fibre-Linked Array-Image Reformatter, Parker \&
Watson (1995)} but upgraded with new gratings, CCD detector and other
refinements. The instrument uses a $1032 \times 1056$~pixel Marconi
CCD47-10 device, with 13$\mu$m pixels. All 6dFGS data taken prior to
October 2002 used 600V and 316R reflection gratings, covering 4000 --
5600\,\AA\ and 5500 -- 8400\,\AA\ respectively. Subsequent data uses
Volume-Phase transmissive Holographic (VPH) 580V and 425R gratings
from Ralcon Development Laboratory, with improved efficiency, focus,
and data uniformity. The wavelength coverage is 3900 -- 5600\,\AA\ and
5400 -- 7500\,\AA\, and the grating and camera angles (and hence
dimensionless resolutions) are identical. The peak system efficiency
(good conditions and acquisition, wavelengths near blaze, good fibres)
is 11\%, but can be much less.

The marginally lower dispersion of the 580V VPH grating, as compared
with the 600V reflection grating, is compensated by the better focus
allowed by the reduced pupil relief. 

The UK Schmidt with 6dF is well-suited to low to medium resolution
spectroscopy of bright ($V<17$), sparsely distributed sources (1 to
50~deg$^{-2}$). As such, it fills the gap left open by 2dF for large,
shallow surveys covering a significant fraction of the total sky. In
terms of $A\Omega$ (telescope aperture $\times$ field of view), UKST/6dF
is similar to AAT/2dF and Sloan.

% ffffffffffffffffffffffffffffffffffffffffffffffffffffffffffffffffffffffffff
\begin{figure}
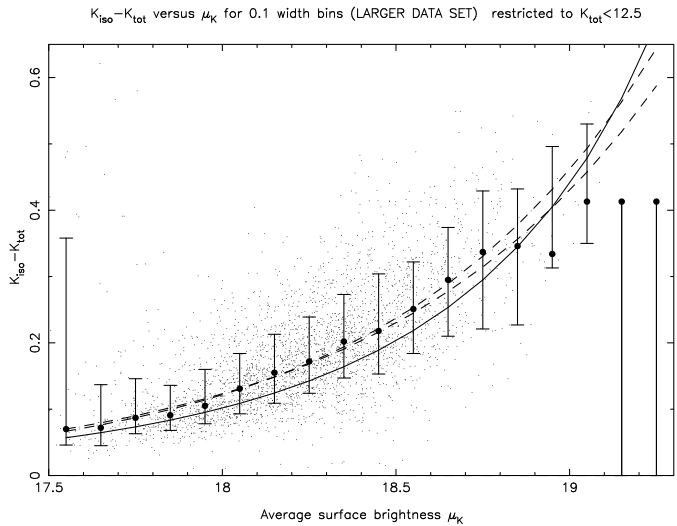

\plotrot{Kisototsb.ps}{0.5}{270} 
\caption{Correlation between isophotal $-$ total magnitude deficit and surface 
brightness, versus a simple exponential disc model ({\em solid curve}) 
and the finally adopted correction ({\em upper dashed curve}). Isophotal
magnitudes are measured to an isophote of $\mu_K=20^m {\rm arcsec}^{-2}$. 
There is increasing Malmquist bias at lower surface brightnesses, since 
galaxies with large deficits are increasingly unlikely to enter the 2MASS 
catalogue.}

\label{fig:Ktotal}
\end{figure}
% ffffffffffffffffffffffffffffffffffffffffffffffffffffffffffffffffffffffffff

% ssssssssssssssssssssssssssssssssssssssssssssssssssssssssssssssssssssss

\section{SURVEY DESIGN}
\label{sec:design}

\subsection{Overview}

Surveys which cover the sky in a new waveband (such as 2MASS), are 
invariably shallow and wide-angle, as this maximises the return (in terms
of sample numbers), for the intrinsically difficult observations. 
This also holds true for the IRAS, ROSAT, HIPASS, NVSS and SUMSS surveys. 
Other projects, such as finding peculiar velocities, are even more strongly 
driven to being as shallow and wide-angled as possible; and any project 
using galaxy distributions to predict dynamics requires just as great sky 
coverage as possible. All such surveys are hence uniquely matched to 6dF, 
with its ability to map the whole sky in realistic timescales.

These arguments have been extended and formalised by Burkey \& Taylor 
(2004), who have recently studied how the scientific returns
of 6dFGS should be optimised in light of existing large-scale datasets
such as the 2dFGRS and SDSS. Their analysis shows that the combined
redshift ($z$) and peculiar velocity ($v$) components of the 6dFGS give
it the power to disentangle the degeneracy between several key
parameters of structure formation, listed in Table~\ref{tab:params}.
They demonstrate that $A_g$, $\Gamma$ and $\beta$ can be determined to
within around 3\% if only the redshift survey is used, although,
$\omega_b$ and $r_g$ are much less well-constrained. If the combined $z$
and $v$ data are used, all of $A_g$, $\Gamma$, $\beta$ and $r_g$ can be
determined to within 2 -- 3\%. The change in $\beta$ and $r_g$ on
different spatial scales can also be determined to within a few percent.
Clearly the advantage of the 6dFGS in understanding structure formation
comes from its large-scale determination of galaxy masses, in addition
to distances.

Burkey \& Taylor also calculate the optimal observing strategy for 6dFGS, 
and confirm that the dense sampling and widest possible areal coverage are 
indeed close to optimal for parameter estimation.

%ttttttttttttttttttttttttttttttttttttttttttttttttttttttttttttttttttttt
\begin{table}
\begin{center}
\caption{Cosmological parameters readily measurable from the 
  6dF Galaxy Survey.
\label{tab:params}
} 
\vspace{6pt}
\begin{tabular}{ll}
\hline \hline
       &        Parameter  \\

\hline
$b$                & bias parameter \\
$A_g = b A_m$      & galaxy power spectrum amplitude \\
$A_v = \Omega_m^{0.6} A_m$  & velocity field amplitude \\
$\Gamma = \Omega_m h$  & power spectrum shape parameter \\
$\omega_b = \Omega_b h$ & mass density in baryons \\
$\beta = \Omega_{m}^{0.6}/b$  & redshift-space distortion parameter \\
$r_g$  & luminous -- dark matter correlation coefficient \\

\hline \hline
\end{tabular}
\end{center}
\end{table}
% ttttttttttttttttttttttttttttttttttttttttttttttttttttttttttttttttttttt

\subsection{Observational Considerations}

The original science drivers for the 6dF project were an
all-southern-sky redshift survey of NIR-selected galaxies, and a large
peculiar velocity survey of early-type galaxies.  Three instrumental
considerations led to the observations for these projects being
merged. Firstly, the combination of the physical size of the 6dF
buttons (5mm), the small plate scale of the Schmidt ($67.14''/$mm),
and the strong angular clustering of the shallow and mostly early-type
input catalogues, meant that acceptable ($\sim 90$\%) completeness
could only be achieved by covering the sky at least twice, in the
sense that the sum of the areas of all tiles observed be at least
twice the actual area of sky covered. Secondly, the spectrograph
optics and CCD dimensions did not simultaneously permit an acceptable
resolution ($R\sim 1000$) over the required minimal wavelength range
4000 -- 7500\,\AA; and this meant that each field had to be observed
separately with two grating setups. Thirdly, the robot configuring
times, and the overheads between fields (parking the telescope,
changing field plates, taking calibration frames, acquiring a new
field), meant that observations less that 1-2 hours/field were not
an efficient use of the telescope. Together, these factors meant that
the redshift survey would necessarily take longer than originally
envisaged. However, careful consideration of the effects of
signal-to-noise and resolution on velocity width measurements (see
Wegner \etal\ 1999), led us to
conclude that for the luminous, high surface brightness galaxies
expected to dominate the peculiar velocity survey, the resolution and
signal-to-noise expected from the redshift survey observations (in
some case repeated to increase S/N), would in general allow velocity
widths to be determined to the required accuracy. Therefore, in early
2001 a decision was made to merge the observations for the two
surveys.

These observational considerations implied that the survey would be of
$\sim 1500$ fields, with $\sim 1$ hour integration time per field per
grating, and covering 4000 -- 7500\,\AA. With 100 -- 135 fibres available
for targets per field, this meant 150 -- 200,000 observations could be
made in total. Given the $\sim 100,000$ targets desired for the
primary $K$-selected survey, and an expected 20\% contingency for
reobservation (either failures or to increase S/N), there remained the
opportunity to include other samples in the survey, especially if they
required lower levels of observational completeness than the primary
sample. Some of these were selected by the Science Advisory Group to
fill out the sample to provide substantial flux-selected samples at
$H,J,I,\bj$, and $\rf$ wavebands; others were invited from the
community as an announcement of opportunity, and resulted in a wide
variety of x-ray, radio, optical, near- and far-infrared selected
extragalactic samples being included. It is striking that most of these 
additional samples derive from the first sky surveys in a new waveband; 
and also that most of them could not be undertaken on any other telescope, 
being too large for long-slit work, but too sparse for multiplexing in 
their own right. 

\subsection{The Primary Sample}
\label{sec:redsurvey}

The primary redshift ($z$-survey) sample is a magnitude-limited
selection drawn from the 2MASS Extended Source Catalog, version 3
(2MASS XSC; Jarrett \etal\ 2000). Since the survey is attempting a
`census' of the local Universe, we want to avoid any bias against
lower-surface-brightness galaxies, and ideally we would use total
magnitudes. 2MASS data does include total magnitudes, estimated from
curves of growth; these are reliable for high galactic latitudes
and/or very bright galaxies, but 2MASS does not have the depth nor
resolution to derive robust total magnitudes for galaxies at lower
latitudes to our desired flux limit. On the other hand, 2MASS includes
very robust isophotal magnitudes ($K_{\rm iso}$) and diameters to an
elliptical isophote of $\mu_K=20^m {\rm arcsec}^{-2}$.  We found that
we were able to make a simple surface-brightness correction to these 
standard isophotal magnitudes, which gave an excellent approximation 
to the total magnitude at high latitudes where they were reliable 
(Fig.~\ref{fig:Ktotal}):
% %%%%%%%%%%%%%%%%%
\begin{equation}
K_{\rm cor} = K_{\rm iso} - 1.5 \exp{1.25 (\overline{\mu_{K20}}-20)} .
\end{equation}
% %%%%%%%%%%%%%%%%%
Here, $\overline{\mu_{K20}}$ is the mean surface brightness within the
$\mu_K=20$ elliptical isophote, and with a maximum allowed correction
of $0.5^m$. This `corrected' isophotal magnitude was also extremely
robust to stellar contamination. There remains a smaller second-order
bias dependent on the convexity of the profile. Further details are in
Burkey (2004).

A latitude cut of $|\,b\,| \geq 10^\circ$ was imposed, mostly because
extinctions closer to the plane would demand much greater intergation
times, and a declination cut of ($\delta < 0^\circ$) was imposed.

Our final selection was then 113\,988 galaxies with 
$K_{\rm cor}<12.75$, corresponding approximately to $K_{20}<13^m$ for 
typical $K$-selected galaxies.

\subsection{The Additional Samples}

Thirteen other smaller extragalactic samples are merged with the primary 
sample. These include secondary 2MASS selections down to $H_{\rm tot} =
13.05$ and $J_{\rm tot}=13.75$ over the same area of sky, constituting an
additional $\sim 5\,000$ sources. 
Optically-selected sources from the SuperCOSMOS
catalogue (Hambly \etal\ 2001) with $\rf\ = 15.6$ and $\bj\ = 16.75$,
$|\,b\,| > 20^\circ$ were included, constituting a further $\sim 20\,000$
galaxies. The remaining miscellaneous piggy-back surveys contribute a
further $\sim 29\,000$ galaxies in various regions of the sky. 
These samples heavily overlap, greatly increasing the efficiency of the
survey - the combined grand sum of all the samples amounts to
500\,000 sources, but these represent only \targets\ different sources
when overlap is taken into account. However, at the current rate of
completion, we estimate that the eventual number of 6dF galaxy redshifts
will be around \aimsixdf.

% ttttttttttttttttttttttttttttttttttttttttttttttttttttttttttttttttttttt
\begin{table*}
\begin{center}
\caption{The 6dFGS target samples used to define the tiling. There are
also samples of 6843 SUMSS sources (Sadler, Sydney) and 466 Durham/UKST
Galaxy Survey sources (Shanks, Durham), not used in the
tiling but included for serendipitious observation. Surveys with
higher priority indices carry greater importance in the allocation of
fields. `Sampling' is as expected from the tiling simulations; in
practice fibre breakages and imperfect fibre assignment reduce these
numbers, especially for lower priority samples.
\label{tab:6dFGStargets}} 
\vspace{6pt}
\begin{tabular}{lccrc}
% \tableline \tableline
\hline \hline
Sample (Contact, Institution)                 & Weight &  Total & Sampling \\
% \tableline
\hline
2MASS $K_s<12.75$ (Jarret, IPAC)       & 8     & 113988 & 94.1\% \\
2MASS $H  <13.05$ (Jarret, IPAC)   & 6     &   3282 & 91.8\% \\
2MASS $J  <13.75$ (Jarret, IPAC) & 6     &   2008 & 92.7\% \\
SuperCOSMOS $r_F<15.6$ (Read, ROE)   2    & 6     &   9199 & 94.9\% \\
SuperCOSMOS $b_J<16.75$ ()Read, ROE   & 6     &   9749 & 93.8\% \\
Shapley             (Proust, Paris-Meudon)   & 6     &    939 & 85.7\% \\
ROSAT All-Sky Survey  (Croom, AAO)  & 6     &   2913 & 91.7\% \\
HIPASS ($>4\sigma$) (Drinkwater, Queensland)    & 6     &    821 & 85.5\% \\
IRAS FSC $6\sigma$ (Saunders, AAO) & 6     &  10707 & 94.9\% \\
DENIS $J<14.00$    (Mamon, IAP)   & 5     &   1505 & 93.2\% \\
DENIS $I<14.85$    (Mamon, IAP)  & 5     &   2017 & 61.7\% \\
2MASS AGN          (Nelson, IPAC)  & 4     &   2132 & 91.7\% \\
Hamburg-ESO Survey  (Witowski,Potsdam) & 4     &   3539 & 90.6\% \\
NRAO-VLA Sky Survey (Gregg, UCDavis)   & 4     &   4334 & 87.6\% \\
% \tableline
\hline
Total                  &       &  167133 & 93.3\% \\
% \tableline \tableline
\hline \hline
\end{tabular}
\end{center}
\end{table*}
% ttttttttttttttttttttttttttttttttttttttttttttttttttttttttttttttttttttt

Table~\ref{tab:6dFGStargets} summarises the breakdown of source
catalogues contributing to the master target list. In total there are
167\,133 objects with field allocations of which two-thirds are 
represented by the near-infrared-selected sample. A further
7\,309 unallocated sources brings the total target list to \targets.
The mean surface density of this
primary sample is $7$~deg$^{-2}$. Literature redshifts have been
incorporated into the redshift catalogue, \zcat\ of these from ZCAT
(Huchra \etal\ 1999) and \twodf\ from the 2000~deg$^{2}$ in common with
the 2dF Galaxy Redshift Survey (Colless \etal\ 2001b).  Roughly half
the sample is early type. For the primary sample, all galaxies are
observed, even where the redshift is already known, to give a complete
spectroscopic sample at reasonable resolution ($R\sim 1000$) and
signal-to-noise ratio (S/N $\sim 10$\perpix). Both tiling (section
3.3) and configuring (3.4) of targets within individual fields used the
weights to assign priorities.

\subsection{Peculiar Velocity Survey}

Peculiar velocities are a vital probe of the large scale mass
distribution in the local universe that does not depend on the
assumption that light traces mass. Early work (Lynden-Bell \etal\ 1988)
made the unexpected discovery of a large ($\sim$600\kms) outflow
(positive peculiar velocities) in the Centaurus region. This led to the
idea of a large extended mass distribution, nicknamed the Great
Attractor (GA), dominating the dynamics of the local universe.
Lynden-Bell \etal\ estimated this structure was located at ($l$, $b$,
$cz$) $\sim$ (307$^\circ$, 7$^\circ$, 4,350\,$\pm$\,350\,km\,s$^{-1}$)
and had a mass of $\sim$5$\times$ 10$^{16}$\,M$_\odot$. Attempts to
measure the expected GA backside infall have proved controversial and
some workers have argued for a continuing high amplitude flow beyond the
GA distance perhaps resulting from a more distant gravitational pull of
the Shapley concentration (312$^\circ$, 31$^\circ$,
14,000\,km\,s$^{-1}$) (Scaramella \etal\ 1989, Hudson \etal\ 1999).

The goal of the peculiar velocity ($v$-survey) is to measure peculiar
velocities for an all-southern-sky sample of galaxies.  Peculiar
velocities are measured for early-type galaxies through the Fundamental
Plane (FP) parameters from 2MASS images and 6dF spectroscopy to give
velocity dispersions. The $v$-survey sample consists of all early-type
galaxies from the primary $z$-survey sample that are sufficiently bright
to yield precise velocity dispersions. Because we cover the sky twice,
suitable candidate galaxies (selected on the basis of either 2MASS
morphology or first-pass 6dF spectroscopy) can be observed a second time
in order to extend the $v$-survey sample to fainter limits. Based on the
high fraction of early-type galaxies in the $K$-selected sample and the
signal-to-noise ratio obtained in our first-pass spectroscopy, we expect
to measure distances and peculiar velocities for 15\,000 galaxies with
$cz < 15\,000$\kms.

When linked with the {\it predicted} peculiar velocities from all-sky
redshift surveys like the PSCz (Branchini \etal\ 1999), a value for
$\Omega$ can be found that is independent of CMB measurements.

\subsection{Field Placement and Tiling Algorithm}
\label{sec:tiling}
The survey area is $17\,046$~deg$^2$, meaning that the 1360 6dF fields
($5.7^\circ$-diameter) contain a mean of 124 sources per field and cover
the sky twice over. An adaptive tiling algorithm was employed to
distribute the fields (``tiles'') across the sky to maximise uniformity
and completeness, described in full in Campbell \etal\ (2004). In brief, 
this consisted of a merit function, which was the priority-weighted sum 
($P=\beta^p$, Sect.~\ref{sec:redsurvey}) of allocated targets; a method 
for rapidly determining fibering conflicts between targets; a method of 
rapidly allocating targets to a given set of tiles so as to maximise 
the merit function; and a method to make large or small perturbations 
to the tiling.
Tiles were initially allocated in random target positions, and the merit 
function maximised via the Metropolis algorithm (Metropolis \etal\ 1953).

It quickly became clear that the clusters were too `greedy' under this 
scheme, in the sense that the completeness was higher in these regions. 
This is easily seen by considering a tiling with a uniform level of 
incompleteness everywhere, but with one last tile still to be placed: 
this will always go to the densest region, as there are the largest density 
of unconfigured targets here also. To counter this effect, we inversely 
weighted each galaxy by the local galaxy surface density (as determined 
from the primary sample) on tile-sized scales; in the above example this 
means the final tile can be placed anywhere with equal merit. This 
achieved our aim of consistent completeness, at a very small penalty in 
overall completeness. It broke down in the heart of the Shapley 
supercluster, with galaxy densities orders of magnitude higher than 
elsewhere, and we added 10 tiles by hand in this region.

\begin{figure}
\plotone{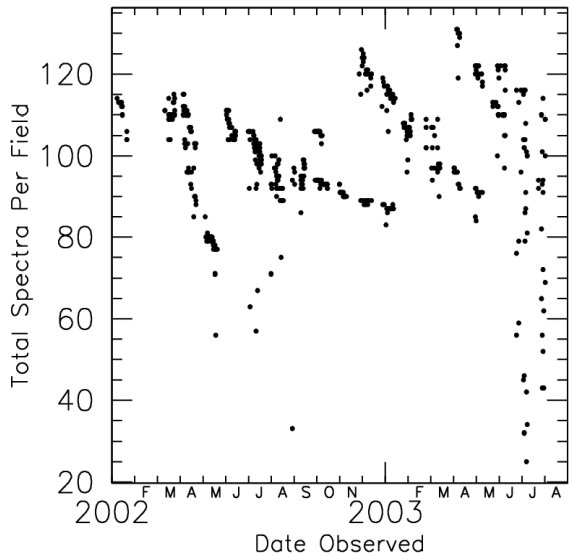}
\caption{Total number of spectra obtained per field
  between January 2002 and June 2003. The upper envelope shows the
  maximum number of fibres available at any one time. It tracks the loss
  of fibres over time, and how Field plate \#1 was taken out of service
  for January-February 2003 for a major fibre repair. The large scatter
  seen from the end of June 2003 is due to a change in observing
  strategy, from field choice maximising fibre allocation to one
  aimed at completing areas of sky.}
\label{fig:attrition}
\end{figure}

Two major tiling runs of the 6dFGS catalogue have been done:
the first in April 2002 before commencement of observations ({\em
  version A}), and a second revised tiling in February 2003 after the
first year of data ({\em version D}). The revision was due to the
higher-than-expected rate at which fibres were broken and temporarily
lost from service (Fig.~\ref{fig:attrition}), and a major revision in 
the primary sample itself from IPAC.

% ffffffffffffffffffffffffffffffffffffffffffffffffffffffffffffffffffffffffff
% ffffffffffffffffffffffffffffffffffffffffffffffffffffffffffffffffffffffffff

\begin{figure*}
\plotfull{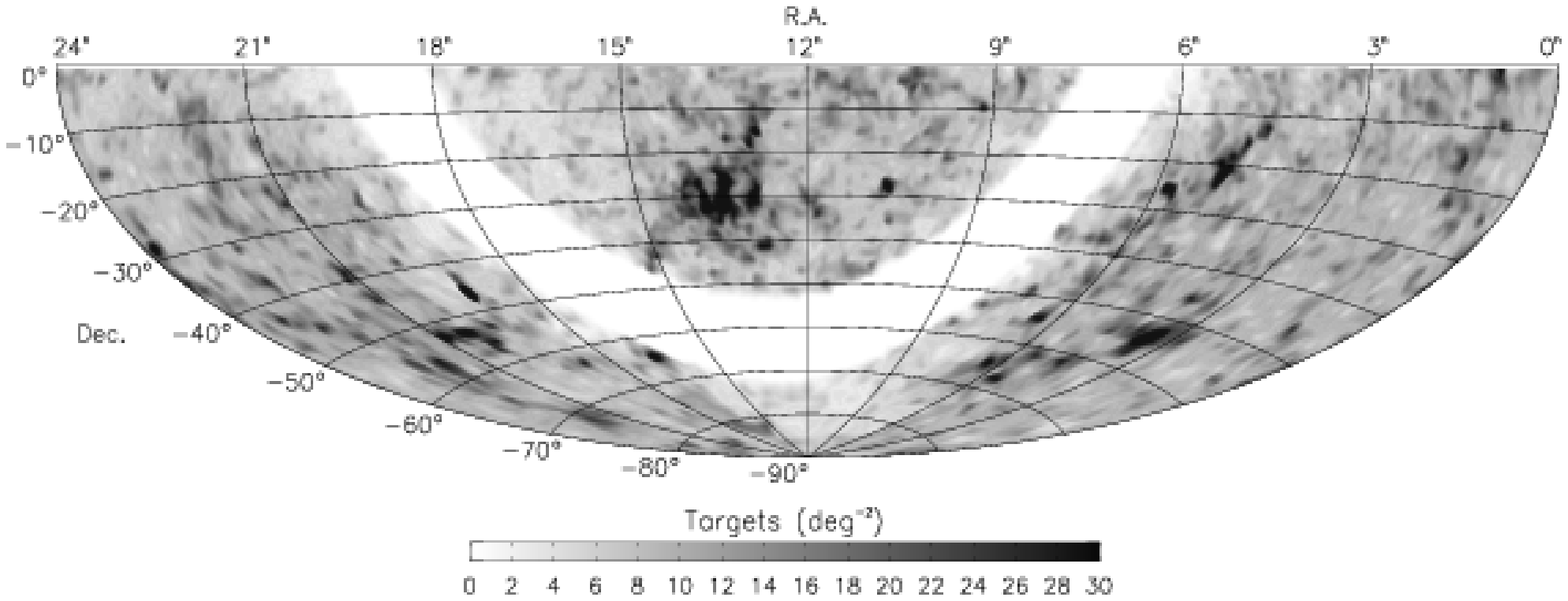}{0.9}
\plotfull{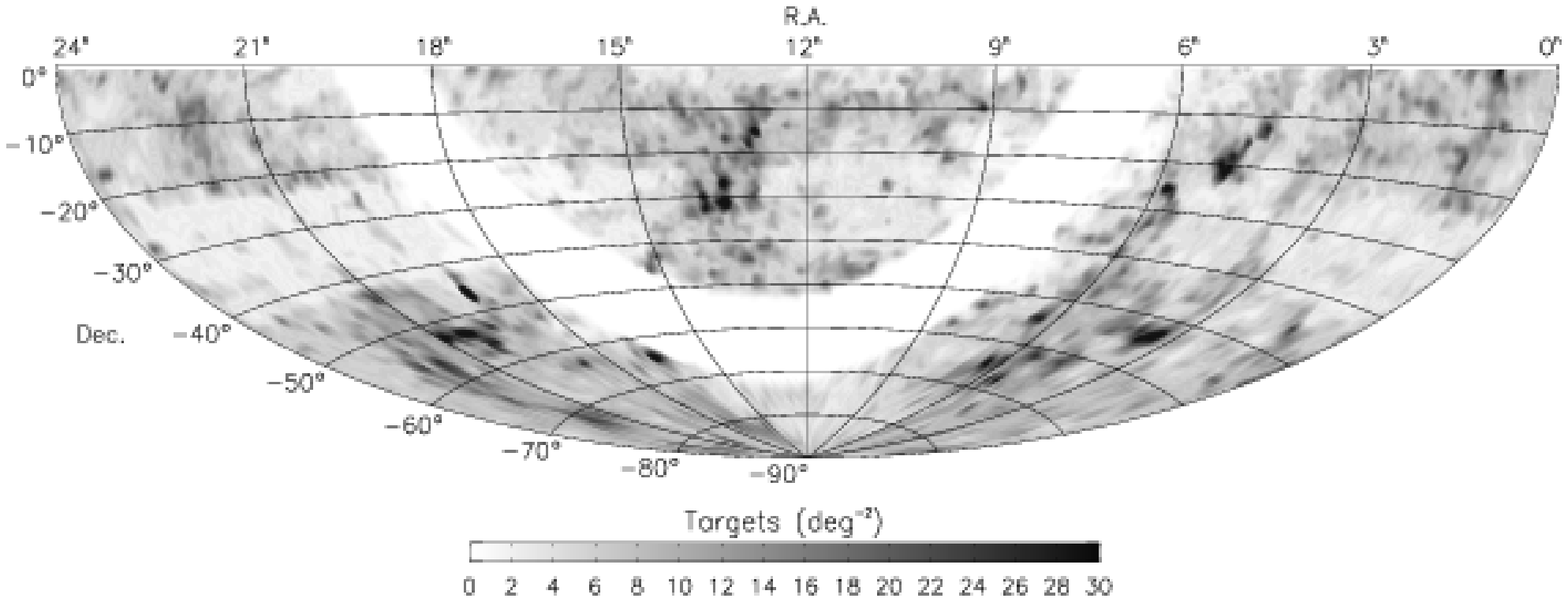}{0.9}
\plotfull{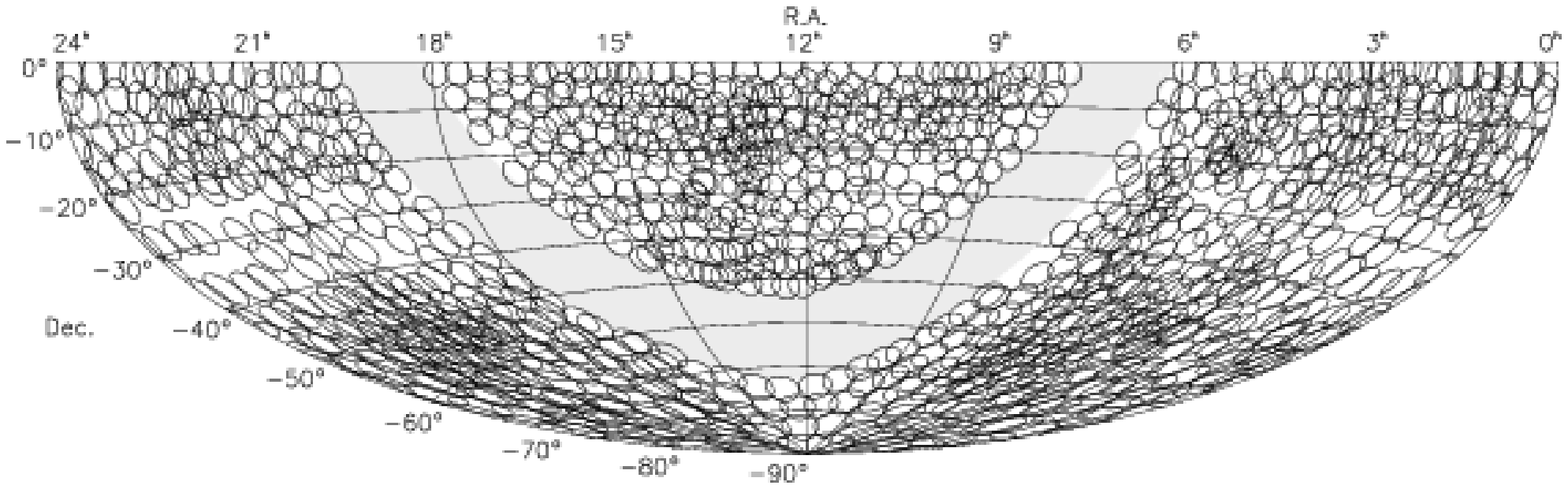}{0.9}
\caption{({\it top}) Sky distribution of all targets in the source 
  catalogues.  ({\it middle}) Distribution of unobserved sources at the
  time of the {\em version D} tiling allocation in February 2003. Note
  how most of the fields observed between the {\em versions A} and 
  {\em D} were confined to the central strip ($- 23^\circ$ to
  $-42^\circ$ declination).  ({\it bottom}) Optimal distribution of
  tiles based on the unobserved sources at the time of {\em version D}.}
\label{fig:tiling}
\end{figure*}

Figure~\ref{fig:tiling} shows the relationships between the full source
list ({\em top}), those that remained unobserved at the time of the 
{\em version D} tiling allocation ({\em middle}), and the optimal tile
placement to cover these ({\em bottom}).

Tests of the two-point correlation function were made on the sample
selected through the final tiling allocation, to see what systematic
effects might arise from its implementation. Mock catalogues were 
generated, with correlation function as observed by the 2dF Galaxy 
Redshift Survey (Hawkins \etal\ 2003), these were tiled as the real 
data and the resulting 2-point correlation function determined and 
compared with the original. This revealed an undersampling on scales 
up to $\sim 1$~\invMpc, clearly the result of the fibre button 
proximity limit. No bias was seen on larger scales.

Theoretical tiling
completenesses of around 95\% were achievable for all except the lowest
priority samples, and variations in uniformity were confined to $< 5$\%. 
However, fibre breakages have meant that 6dFGS has consistently run with 
many fewer fibres than anticipated, impacting on the completeness of the 
lower priority samples in particular. With a fixed timeline for the 
survey (mid 2005) and a fixed number of fields to observe, there is 
little choice in the matter.

% ffffffffffffffffffffffffffffffffffffffffffffffffffffffffffffffffffffffffff
% ffffffffffffffffffffffffffffffffffffffffffffffffffffffffffffffffffffffffff

\subsection{Fibre Assignment}

Within each tile, targets are assigned to fibres by the same {\tt
  CONFIGURE} software used by the 2dF Galaxy Redshift Survey.  This
iteratively tries to find the largest number of targets assigned to
fibres, and the highest priority targets.  Early configurations (until
mid-2003) were usually tweaked by hand to improve target yield, after
that date a revised version of {\tt CONFIGURE} was installed with much
improved yields and little or no further tweaking was in general made.

% ssssssssssssssssssssssssssssssssssssssssssssssssssssssssssssssssssssss

\section{SURVEY IMPLEMENTATION}
\label{sec:implementation}

\subsection{Observational Technique}

Field acquisition with 6dF is carried out using conventional guide-fibre
bundles. Four fibre buttons are fitted with coherent bundles of seven
fibres rather than a single science fibre. Fibre diameter is 100$\mu$m
(6.7\arcsec) and the guide fibres are in contact at the outer cladding
to give a compact configuration $\sim$20 arcsec in diameter. These
fibres are 5~m long and feed the intensified CCD acquisition camera
of the telescope. The use of acquisition fibres of the same diameter as the
science fibres is sub-optimal, but in practice the four guide-fibre
bundles give good alignment, particularly as guide stars near the edge
of the field are always chosen.

Guide stars are selected from the Tycho-2 catalogue (Hoeg \etal\ 2000),
and have magnitudes typically in the range $8<V<11$. Field acquisition
is straightforward in practice, and the distortion modelling of the
telescope's focal surface is sufficiently good that a field rotation
adjustment is not usually required, other than a small standard offset.

Each field is observed with both V and R gratings, these are later
spliced to reconstruct a single spectrum from these two observations.
Integrations are a minimum of 1~hr for the V spectrum and 0.5~hr for
the R spectrum, although these times are increased in poor observing
conditions. This gives spectra with typical S/N around 10~pixel$^{-1}$,
yielding $>$90\% redshift completeness.

This observing strategy typically allows 3-5 survey fields to be
observed on a clear night, depending on season. With 75\% of the UKST
time assigned to 6dFGS, and an average clear fraction of 60\%, we
typically observe about 400 fields per year.  The observational strategy
is to divide the sky into three declination strips. Initially, the
survey has concentrated on the $\delta=-30^\circ$ declination strip
(actually $-42^\circ<\delta<-23^\circ$); the equatorial strip
($-23^\circ<\delta<0^\circ$) will be done next, and then finally the
polar cap ($\delta<-42^\circ$).

Observations started in June 2001, though final input catalogues and
viable reduction tools were not available until 2002. Early data suffered
from various problems, including poorer spectrograph focus due to
misalignment within the camera; poorer quality control; and use of
preliminary versions of the 2MASS data, leading to many observed sources
being dropped from the final sample. The 2001 data are not included in
this data release.

Initial observations were carried out at mid-latitudes for observational
convenience, with the actual band corresponding to one of the Additional
Target samples. Excursions from this band were made to target other
Additional Target areas, where separate telescope time had been allotted
to such a program, but the observations could be fruitfully folded in to
6dFGS.

The observing sequence conventionally begins with R data (to allow a
start to be made in evening twilight). With the telescope at access
park position, a full-aperture flat-field screen is illuminated with
calibration lamps. First of these is a set of quartz lamps to give a
continuum in each fibre. This serves two purposes; (a) the loci of the
150 spectra are defined on the CCD, and (b) the differences between
the extracted spectra of the smooth blackbody lamp allow flatfielding
of the signatures introduced into the object spectra by pixel-to-pixel
variations and fibre-fibre chromatic throughput variations. Then the
wavelength-calibration lamps are exposed, HgCd + Ne for the R data and
HgCd + He for the V data. After the R calibration exposure, the field
is acquired and the 3$\times$10-min red frames obtained. Once they are
completed, the grating is changed remotely from the control room and
the 3$\times$20-min V frames obtained.  At the end of the sequence,
the V wavelength calibration and flat-field exposures are made.

With the change of field comes a change of slit-unit (because of the two
6dF field plates), so all the calibrations must be repeated for the next
field. Usually, the reverse waveband sequence is followed, i.e.,
beginning with V and ending with R. This process continues throughout
the night, as conditions allow.

\subsection{Reduction of Spectra}
\label{sec:reduction}

The reduction of the spectra uses a modified version of the {\tt 2DFDR}
package developed for the 2dF Galaxy Redshift Survey. Unlike 2dF data,
tramline fitting is done completely automatically, using the known gaps
in the fibres to uniquely identify the spectra with their fibre number.
Because of computing limitations, {\tt TRAM} rather than {\tt FIT}
extractions are performed. {\tt FIT} extractions would reduce crosstalk
between fibres, but this is already small for 6dF compared with 2dF.
Scattered light subtraction is not in general performed, unless there is
specific reason for concern, such as during periodic oil-contamination
episodes within the dewar. Again, scattered light performance is better
with 6dF in general than with 2dF.

% ffffffffffffffffffffffffffffffffffffffffffffffffffffffffffffffffffffffffff
\begin{figure*}
\plotfull{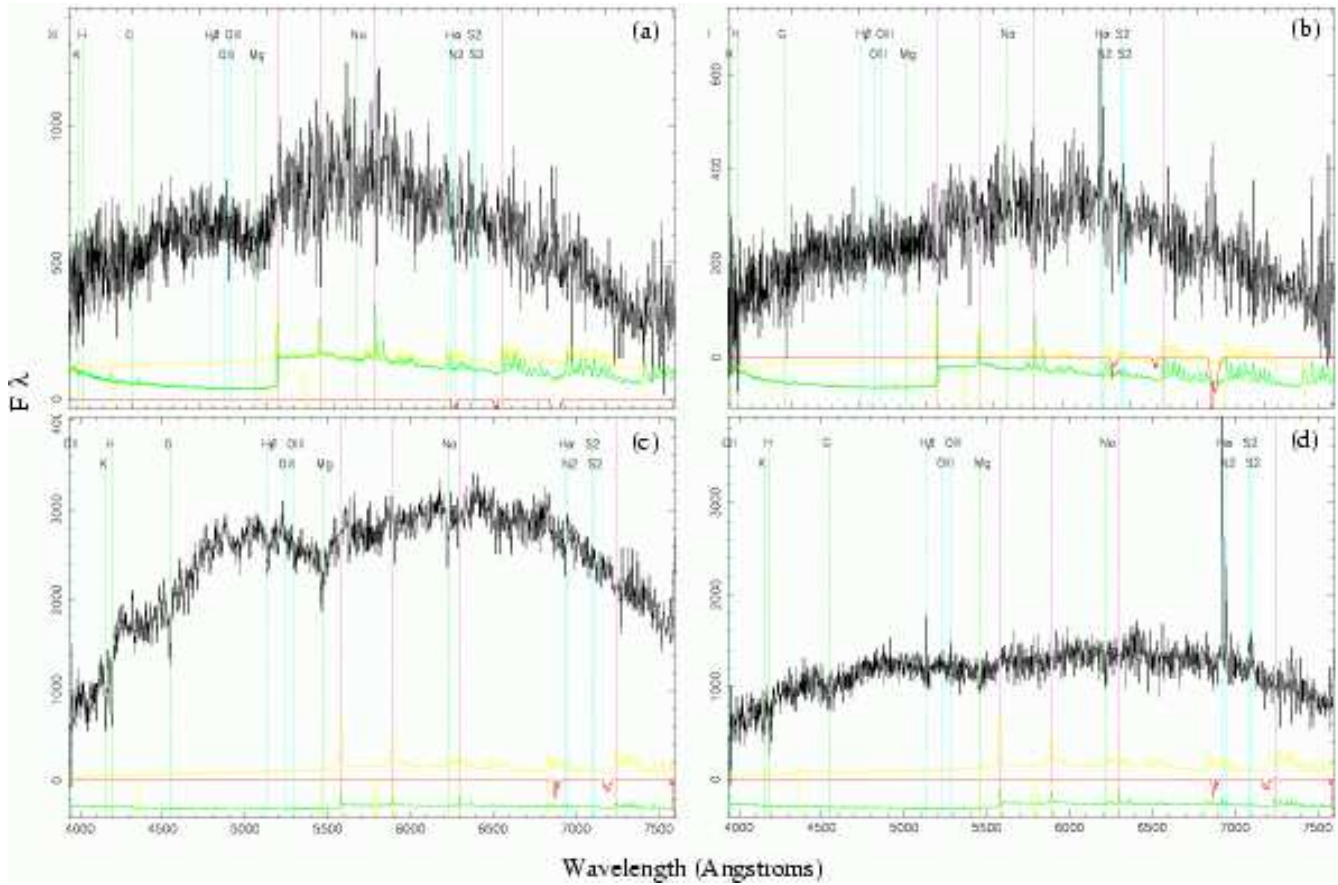}{1.0}
\caption{Examples of galaxy spectra from the 6dFGS exhibiting a range
  of types and redshift quality $Q$:
($a$) $Q=3$ ABS at $z=0.04601$. ($b$) $Q=3$ EMI at $z=0.03377$.
($c$) $Q=4$ ABS at $z=0.05645$. ($d$) $Q=4$ EMI at $z=0.05344$.  
Here, EMI refers to emission-line redshifts (and qualities), 
and ABS refers to absorption line determinations.
Major absorption and emission-line features in the spectra have been
labelled.}
\label{fig:examplespect}
\end{figure*}
% ffffffffffffffffffffffffffffffffffffffffffffffffffffffffffffffffffffffffff

The extracted spectra for each field are combined, usually weighted by
S/N to cope with variable conditions. The S/N is computed at this stage,
and a S/N per pixel of 10 in each of V and R frames usually indicates a
satisfactorily observed field. All data are then fluxed using 6dF
observations of the standard stars Feige 110 and EG274. This fluxing is
inevitably crude, in that the same fixed average spectral transfer
function is assumed for each plate for all time. Differences in the
transfer function between individual fibres are corrected for by
the flat-fielding.

The resulting R and V spectra for each source are then spliced together,
using the overlapping region to match their relative scaling. In order
to avoid a dispersion discontinuity at the join in each spectrum, we
also rescrunch the lower dispersion R data onto an exact continuation of
the V wavelength dispersion.

\subsection{Spectral Quality}

Most spectra have no problems, in the sense that: (1) the S/N is reasonable
given the magnitude of the source; (2) both V and R frames are
available; and (3) there were no problems in the reduction. However,
there are significant caveats of which all users should be aware.

\begin{itemize}
\item{Many fields were observed in marginal conditions, and have reduced
overall S/N as a result. Our philosophy has been to extract what good
spectra we can from these fields, and recycle the rest for
reobservation. A field was only reobserved in its entirety where the
data was valueless.}

\item{Many fields were observed with three or occasionally even two guide
fibres, with consequent lower and more variable S/N.}

\item{Some fibres have poor throughputs due to misalignment or poor
glueing within the button, and variations of factors of two are normal.}

\item{Many fibres, throughout the duration of the survey, have suffered 
various damage in use, short of breakage. Very often, this has resulted in
strong fringing in the spectral response of the fibre, due
to an internal fracture acting as a Fabry-Perot filter. This
did not often flat-field out completely.}

\item{The CCD is in any case a thinned blue-sensitive chip; as a result,
red data suffers increasing levels of fringing towards longer wavelengths, 
and this does not always flat-field out.}

\item{Fibre breakages during configuring, or between blue and red
observations, or severe differences in acquisition, can lead to occasional
missing or mis-spliced red or blue data.}

\item{Some fields have missing red data.}

\item{Though scattered light is not a major problem in general, data
at the blue end of the spectra can be corrupted, because the actual
counts are so low. In extreme cases, the spectra can become
negative. The overall quality of the fluxing is untested, and should
be treated with extreme caution.}

\item{All VPH data suffer from a faint but variable, spurious,
spectral feature at wavelengths around 4440\,\AA\ (10 pixel region) 
in the V grating, and 6430 and 6470\,\AA\ (10 pixel regions) in the R. 
The reason, after extensive investigation, was
determined to be a ghost caused by dispersed light reflected back off
the grating and recollimated by the camera, being undispersed in first
order reflection mode by the VPH grating, and refocused onto the chip
as a somewhat out-of-focus (10-20 pixel diameter), undispersed, image
of the fibre, with an intensity 0.1-1\% of the summed dispersed
light. Circumventing this problem requires tilting the fringes within
the grating (so they are no longer parallel with the normal to the
grating) by a degree or two, to throw the ghost image just off the
chip.}

\end{itemize}

\subsection{Redshift Measurement}
\label{sec:redshifts}

Accurate redshift measurement is a fundamental componenent of both the
$z$ and $v$-surveys. We started with the semi-automated redshifting {\tt
RUNZ} software used for the 2dF Galaxy Redshift Survey (Colless \etal\ 
2001b), kindly provided by Will Sutherland. Extensive modifications
were made in order to accept 6dF data. The version used for 2dF
determined quasi-independent estimators of the redshift from emission
and absorption features; this improved the reliability of the redshift
estimates, while reducing their accuracy. Since the line identication of
the higher S/N and higher dispersion 6dF spectra was usually not in
doubt, we decided in general not to patch out emission features in
determining cross-correlation redshifts; and in general the
cross-correlation redshift was used in preference to the emission-line
redshift.

% ffffffffffffffffffffffffffffffffffffffffffffffffffffffffffffffffffffffffff
\begin{figure}
% \dummyfigure
\plotone{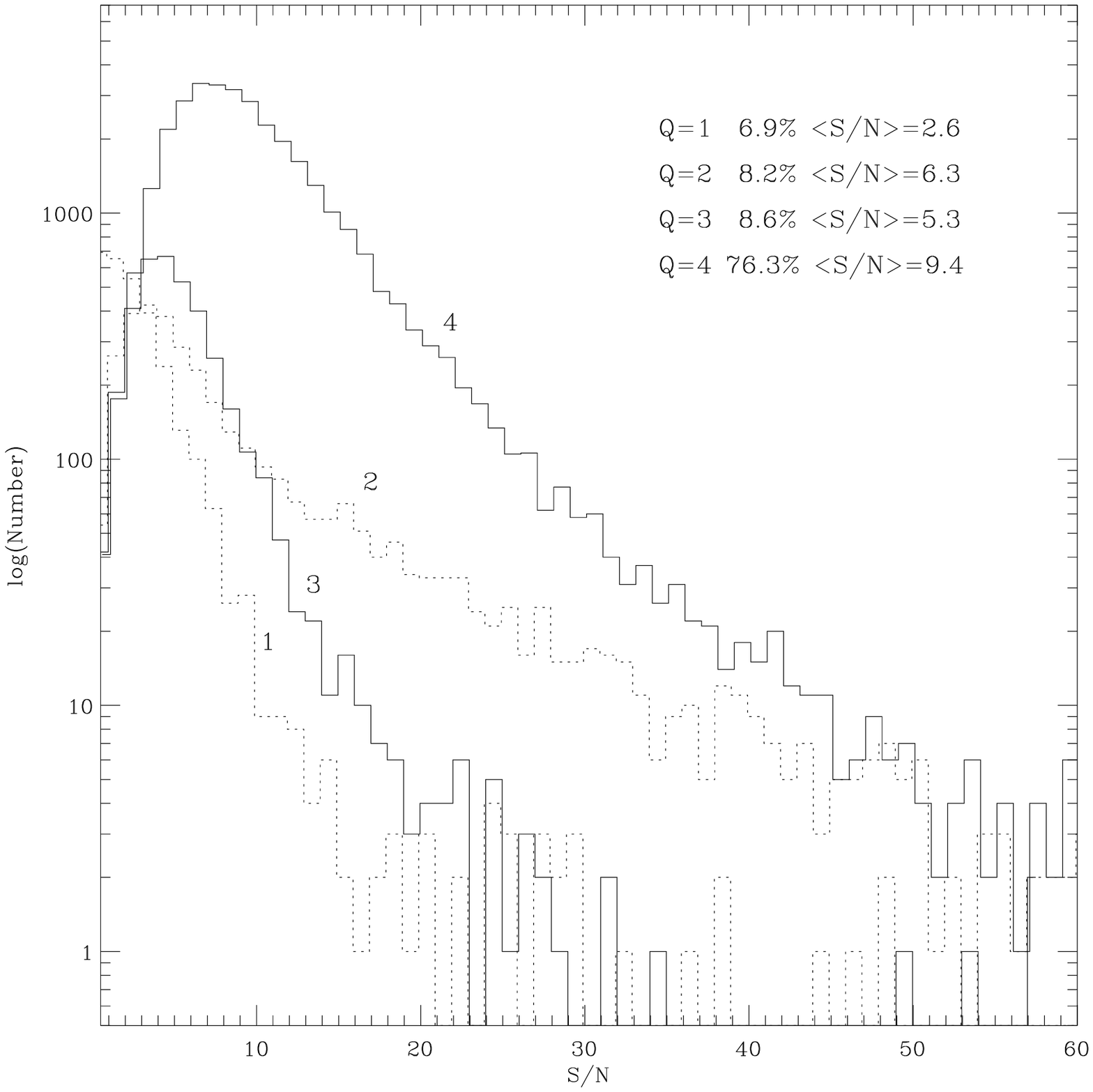}
\caption{Distribution of signal-to-noise for each class of redshift
  quality ($Q = 1$ to 4) for the 6dFGS. The $Q=5$ sources are too few
  (24) to show.}
\label{fig:qualitySN}
\end{figure}
% ffffffffffffffffffffffffffffffffffffffffffffffffffffffffffffffffffffffffff

Each automated redshift is checked visually to decide whether the
software has made an accurate estimate or been misled by spurious
spectral features. Such features are typically due to fibre interference
patterns or poor sky subtraction and are difficult to identify through
software, although easily recognisable to a human operator.  The
operator checks the automated redshift by comparing it to the original
spectrum, the location of night-sky line features and the
cross-correlation peak. In some cases, manual intervention in the form
of re-fitting of spectral features or of the correlation peaks makes for
a new redshift. In the majority of cases, however, the automated
redshift value is accepted without change. The final redshift value is
assigned a quality, $Q$, between 1 to 5 where $Q=3,4,5$ for redshifts
included in the final catalogue. $Q=4$ represents a reliable redshift
while $Q=3$ is assigned to probable redshifts; $Q=2$ is reserved for
tentative redshift values and $Q=1$ for spectra of no value. $Q=5$
signifies a `textbook' high signal-to-noise spectrum,
although in practice is used rarely for the 6dFGS. 
Figure~\ref{fig:examplespect}
shows a few examples of galaxy spectra across the range of redshift
quality, for both emission and absorption-line spectra.

The same visual assessment technique was employed for the 2dF Galaxy
Redshift Survey and greatly increased the reliability of the final
sample: repeat measurements on a set of $\sim 15\,000$ 2dF spectra by two
operators were discrepant in only 0.4\% of cases (Colless \etal\
2001b).  Figure~\ref{fig:qualitySN} shows the relationship between
redshift quality $Q$ and mean signal-to-noise of the spectra that
yielded them.  The vast majority (76\%) of the 6dFGS redshifts have
$Q=4$ from spectra with a median signal-noise of 9.4~\perpix. The
$Q=5$ sources are too few (24) to show. For $Q=3$ redshifts the median
signal-to-noise ratio drops to 5.3~\perpix, indicating minimum of
range of redshift-yielding spectra. In both cases, note the long tail
to higher signal-to-noise values. The median signal-to-noise ratio for
$Q=2$ redshifts (6.3~\perpix) is slightly higher than that for $Q=3$
(5.3~\perpix). This is due to the significant number of Galactic
sources such as stars and planetary nebulae, which produce high
signal-to-noise spectra, but are assigned $Q=2$ on account of their
zero redshift.

% ffffffffffffffffffffffffffffffffffffffffffffffffffffffffffffffffffffffffff
\begin{figure*}
\plotfull{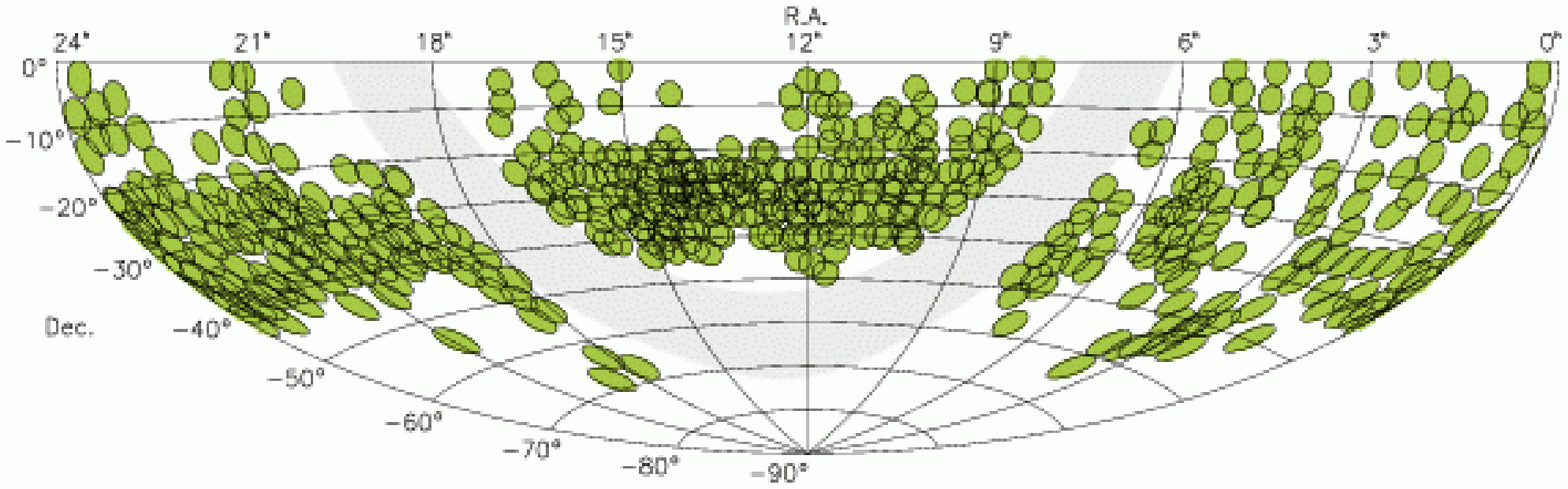}{0.9}
\plotfull{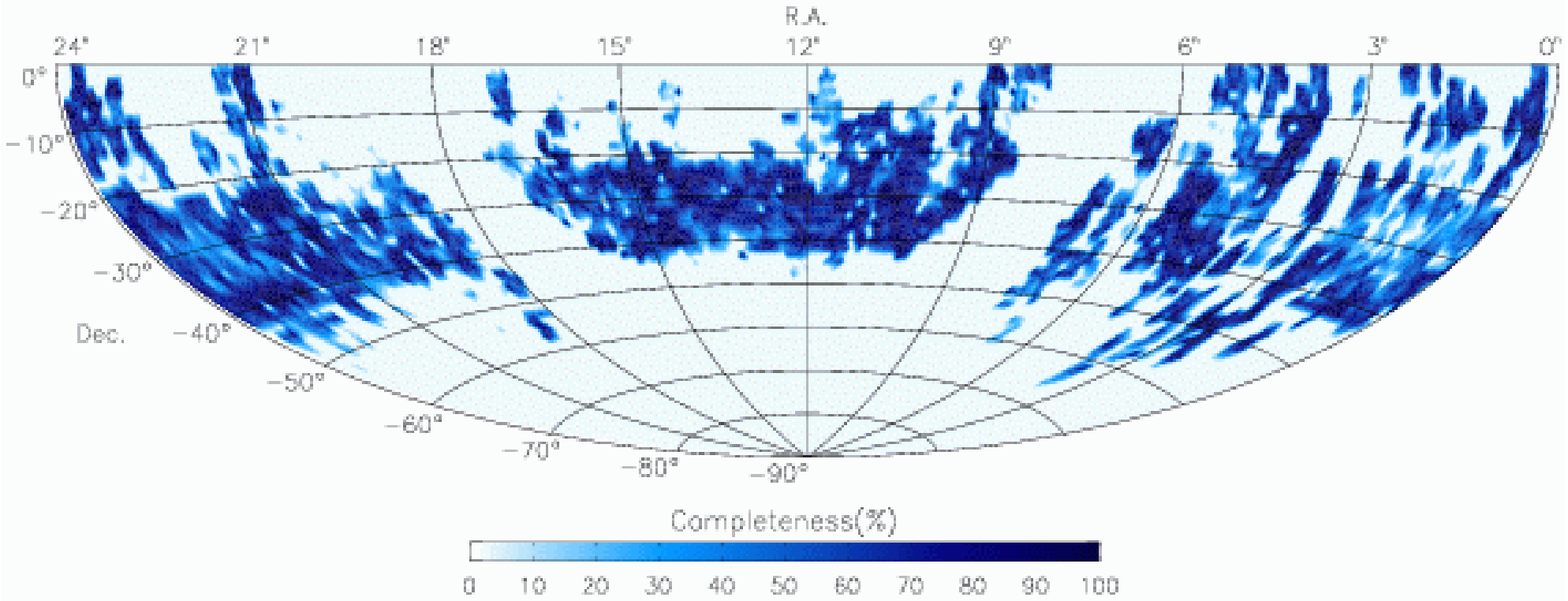}{0.9}
\caption{({\em top}) Location of the observed fields contributing redshifts to 
  the First Data Release. ({\em bottom}) Redshift completeness on the
  sky, combining the 6dF First Data Release redshifts with the
  literature sources.}
\label{fig:obsflds}
\end{figure*}
% ffffffffffffffffffffffffffffffffffffffffffffffffffffffffffffffffffffffffff

% ssssssssssssssssssssssssssssssssssssssssssssssssssssssssssssssssssssss

\section{FIRST DATA RELEASE}
\label{sec:datarelease}

\subsection{Statistics and Plots}

Between January 2002 and July 2003 the 6dF Galaxy Survey Database compiled
\totalDB\ spectra from which \uniqueDB\ unique redshifts were derived.
The numbers of spectra with redshift quality $Q \ge 3$ were \totalDBqual\ 
for the full set and \uniqueDBqual\ for the unique redshifts.  
Of the \targets\ total galaxies in the target sample, \lit\ had
existing literature redshifts: \zcat\ from the ZCAT compilation (Huchra
\etal\ 1999) and \twodf\ from the 2dF Galaxy Survey (Colless \etal\ 2001b).
Of the 113\,988 $K_s$-selected sources, there are 32\,156 6dF-measured
redshifts of redshift quality $Q \ge 3$, plus a further 21\,151
existing literature redshifts. 
Table~\ref{tab:breakdown} summarises these values for the individual
sub-samples as they appear in the 6dFGS Database.

% ttttttttttttttttttttttttttttttttttttttttttttttttttttttttttttttttttttt
\begin{table*}
\begin{center}
\caption{Status of the 6dFGS target samples, as listed in the
  database.
\label{tab:breakdown}
}
\vspace{6pt}
\begin{tabular}{rlccccccccccccc}
% \tableline \tableline
\hline \hline

   id & survey &total &$cz\le600$ &$cz>600$ &6df $z$ &lit $z$ &6df$>600$
 &Q345 &Q1 &Q2 &Q3 &Q4 &Q5 &no $z$ \\
 \hline \hline

    1 &2MASS $K_s<12.75$      & 113988 & 1750 & 53051 & 33650 & 21151 & 32983 & 32156 & 1312 & 1494 & 2708 & 29433 & 15 & 59187 \\
    3 &2MASS $H  <13.05$      & 3282 & 18 & 853 & 526 & 345 & 512 & 492 & 33 & 34 & 58 & 434 & 0 & 2411 \\
    4 &2MASS $J  <13.75$      &2008 & 17 & 552 & 333 & 236 & 319 & 304 & 14 & 29 & 28 & 276 & 0 & 1439 \\
    5 &DENIS $J  <14.00$      & 1505 & 11 & 259 & 124 & 146 & 117 & 111 & 26 & 13 & 27 & 84 & 0 & 1235 \\
    6 &DENIS $I<14.85$        & 2017 & 96 & 191 & 150 & 137 & 63 & 63 & 18 & 87 & 10 & 53 & 0 & 1730 \\
    7 &SuperCOSMOS $r_F<15.6$ & 9199 & 137 & 3310 & 1539 & 1908 & 1439 & 1407 & 46 & 132 & 104 & 1302 & 1 & 5752 \\
    8 &SuperCOSMOS $b_J<16.75$&  9749 & 35 & 3718 & 1973 & 1780 & 1961 & 1900 & 76 & 73 & 173 & 1726 & 1 & 5996 \\
   78 &Durham/UKST extension  &  466 & 2 & 73 & 10 & 65 & 8 & 8 & 1 & 2 & 6 & 2 & 0 & 391 \\
   90 &Shapley supercluster   & 939 & 9 & 323 & 282 & 50 & 273 & 250 & 22 & 32 & 48 & 202 & 0 & 607 \\
  113 &ROSAT All-Sky Survey   & 2913 & 99 & 535 & 395 & 239 & 300 & 223 & 231 & 172 & 53 & 170 & 0 & 2279 \\
  116 &2MASS red AGN Survey   & 2132 & 9 & 252 & 129 & 132 & 121 & 81 & 106 & 48 & 45 & 36 & 0 & 1871 \\
  119 &HIPASS ($>4\sigma$)    & 821 & 8 & 268 & 135 & 141 & 130 & 121 & 11 & 14 & 29 & 92 & 0 & 545 \\
  125 &SUMSS/NVSS radio sources   & 6843 & 321 & 709 & 654 & 376 & 347 & 322 & 89 & 332 & 51 & 270 & 1 & 5813 \\
  126 &IRAS FSC ($6\sigma$)   & 10707 & 258 & 2872 & 1360 & 1770 & 1218 & 1105 & 303 & 255 & 198 & 906 & 1 & 7577 \\
  129 &Hamburg-ESO Survey QSOs&  3539 & 73 & 197 & 220 & 50 & 150 & 56 & 204 & 164 & 19 & 37 & 0 & 3269 \\
  130 &NRAO-VLA Sky Surv. QSOs&  4334 & 342 & 146 & 483 & 5 & 142 & 62 & 303 & 421 & 42 & 20 & 0 & 3846 \\
         & & & & & & & & & & & & & & \\
      & Total  & 174442 & 3185 & 67309 & 41963 & 28531 & 40083 & 38661 & 2795 & 3302 & 3599 & 35043 & 19 & 103948 \\
 \hline \hline

\end{tabular}
\end{center}
\begin{flushleft}
{\bf Column Headings:}\\
% \begin{array}
$cz\le600$ --- object has a redshift (either 6dF-measured with quality $>
1$ or from the literature) less than or equal to 600\kms. \\
$cz>600$ --- object has a redshift (either 6dF-measured with quality $> 1$
or from the literature) greater than 600\kms.\\
6df $z$ --- total number of 6dF-measured redshifts with quality $Q > 1$.\\
lit $z$ --- total number of literature redshifts.\\
6df$>600$ --- number of 6dF-measured redshifts greater than 600\kms with
quality $Q > 1$.\\
Q345 --- total number of (6dF-measured) sources with redshift quality $Q =
3, 4$ or 5.\\
Q1, Q2, Q3, ... --- total number of sources with redshift quality $Q = 1$,
$Q=2$, $Q=3$, etc.\\
no $z$ --- number of sources in the database with neither a 6dF (quality
$Q > 1$) nor literature redshift.\\
\end{flushleft}
\end{table*}
% ttttttttttttttttttttttttttttttttttttttttttttttttttttttttttttttttttttt

Data from \fields\ fields have contributed to the first data release. As
shown in Fig.~\ref{fig:obsflds}\,({\em top}), the majority of these occupy
the central declination strip between $-42^\circ < \delta < 23^\circ$.
Overall there are \totaltiles\ on the sky: \tilesA\ in the equatorial
strip, \tilesB\ in the central strip, and \tilesC\ in the polar region. 
Figure~\ref{fig:obsflds}\,({\em bottom}) shows the corresponding
distribution of redshift completeness on the sky for the $K$-band
sample. The {\sl redshift completeness}, $R$, is that fraction of
galaxies in the parent catalogue of \targets\ with acceptable ($Q\ge3$)
redshifts in a given area of sky, from whatever source,
% eeeeeeeeeeeeeeeeeeeeeeeeeeeeeeeeeeee
% \begin{equation}
\begin{eqnarray}
R &  = & \frac{N_z(\btheta)}{N_p(\btheta)}  \nonumber\\
           &   = & \frac{ N_{{\rm lit}}(\btheta) + N_{{\rm 6dF}}(\btheta) }
      { N_{{\rm lit}}(\btheta) + N_{{\rm 6dF}}(\btheta) +
 N_{\rm Gal}(\btheta) + N_{\rm f}(\btheta) +
N_{\rm r}(\btheta)}
\label{redshiftcompl}
\end{eqnarray}
% \end{equation}
% eeeeeeeeeeeeeeeeeeeeeeeeeeeeeeeeeeee
Here, $N_p(\btheta)$ is the number of galaxies from the parent
catalogue (per unit sky area) at the location $\btheta$, and
$N_z(\btheta)$ is the number with redshifts, either from 6dF ($N_{{\rm
6dF}}(\btheta)$) or the literature ($N_{{\rm lit}}(\btheta)$). Sources
in the parent catalogue that have been redshifted and excluded are
either stars, planetary nebulae/ISM features (both assigned $Q=2$), or
failed spectra ($Q=1$).  In Eqn.~\ref{redshiftcompl} their numbers are
denoted by $N_{\rm Gal}(\btheta)$ and $N_{\rm f}(\btheta)$. The
remaining sources are those yet to be observed, $N_{\rm r}(\btheta)$. 
Of the first $\sim 41\,000$ sources observed with 6dF,
around 3\% were stars, 1\% were other Galactic sources, and 11\%
failed to yield a redshift.

\begin{figure}
\plotone{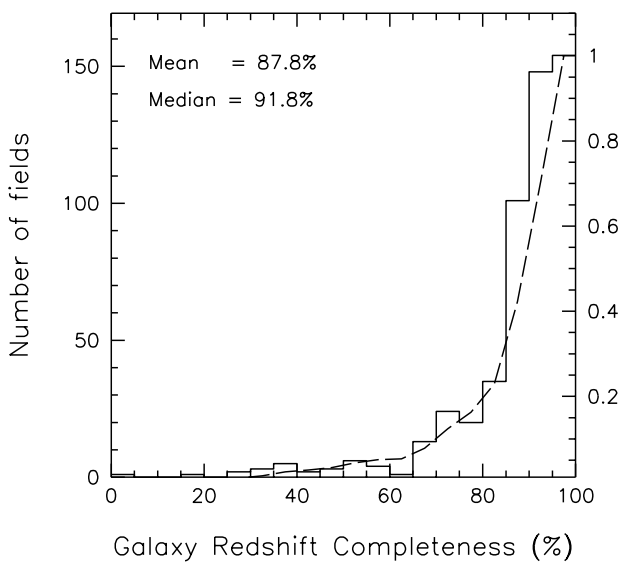}
\caption{Galaxy redshift completeness by field, where completeness
  is the number of 6dF redshifts over the total 6dF redshifts and
  failures. The dashed line indicates the cumulative fraction according to 
  the right-hand axis.}
\label{fig:fldcompl}
\end{figure}

The {\sl field completeness} is the ratio of acceptable redshifts in a
given field to initial sources, and hence is only relevant to targets
observed with 6dF. It also excludes Galactic features like stars and
ISM. Figure~\ref{fig:fldcompl} shows the distribution of field
completeness from the first \fields\ fields and its cumulate. This
demonstrates that the redshift success rate of 6dF is good, with
both the median and mean completeness around 90\%.

% ffffffffffffffffffffffffffffffffffffffffffffffffffffffffffffffffffffffffff
% ffffffffffffffffffffffffffffffffffffffffffffffffffffffffffffffffffffffffff

Observe the large difference between the high {\sl field} completeness
values of Fig.~\ref{fig:fldcompl} and the lower {\sl redshift}
completeness in Fig.~\ref{fig:obsflds}\,({\em bottom}). This is due to the
high degree of overlap in the 6dFGS field allocation. The large variance
in the density of targets has meant that most parts of the sky need to
be tiled two or more times over. This is not at all obvious in
Fig.~\ref{fig:obsflds}\,({\em top}) which superimposes all fields, giving
the impression of a single layer of tiles. While much of the central
strip contains observed and redshifted fields, it also contains other
fields in this same region, as yet unobserved.

\begin{figure}
\plotone{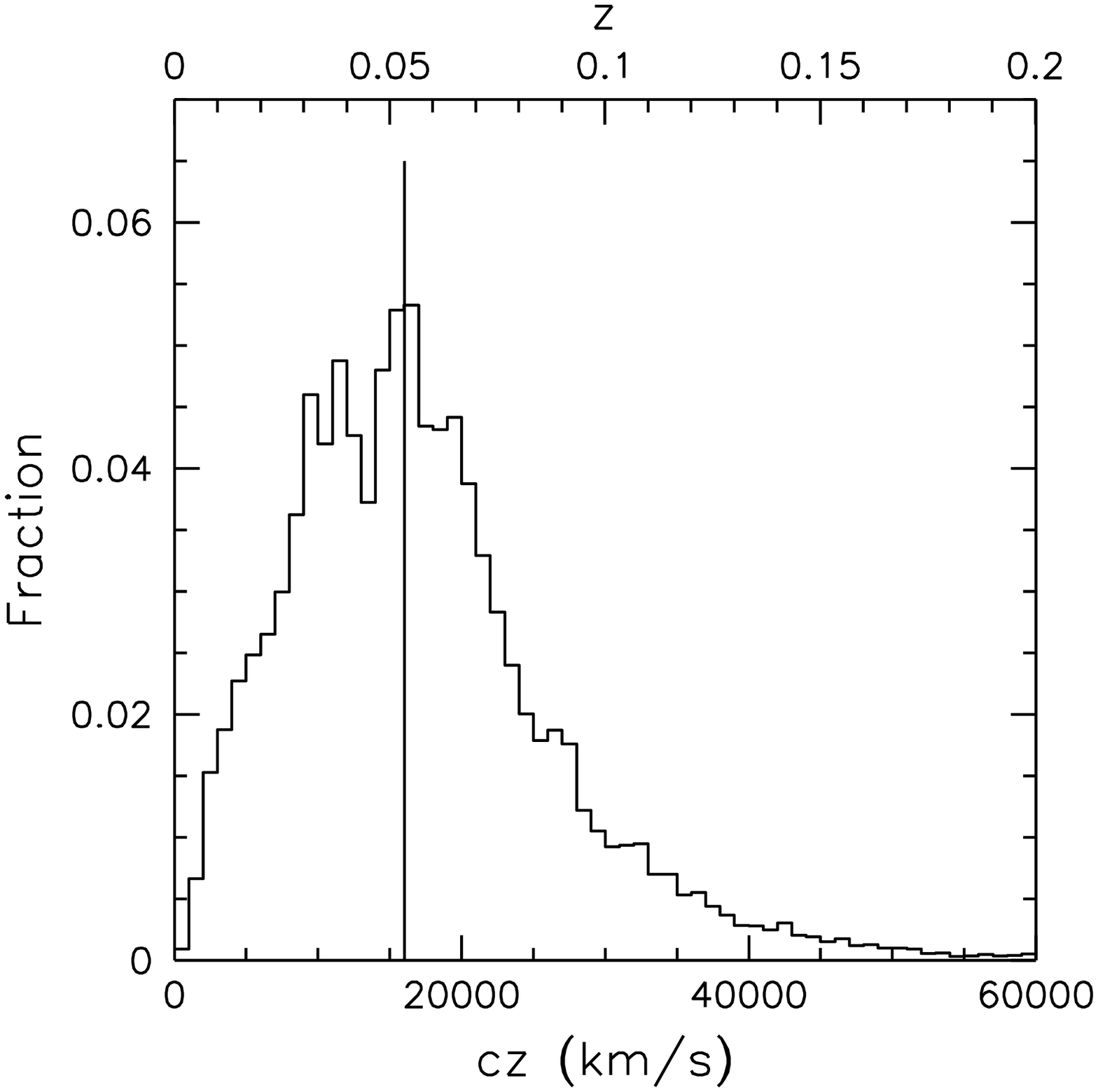}
\caption{Distribution of redshifts for the 6dF First Data
  Release galaxies with redshift quality $Q \ge 3$ and $cz > 600$~\kms.
  The mean redshift for the survey ($\langle cz \rangle = 16\,008$~\kms)
  is indicated with a vertical solid line.}
\label{fig:nz}
\end{figure}

The distribution of 6dFGS redshifts exhibits the classic shape for
magnitude-limited surveys of this kind (Fig.~\ref{fig:nz}). The median
survey redshift, $\langle cz \rangle = 16\,008$~\kms\ ($\bar{z} =
0.055$), is less than half that of the 2dFGRS or SDSS surveys.

% ffffffffffffffffffffffffffffffffffffffffffffffffffffffffffffffffffffffffff
% ffffffffffffffffffffffffffffffffffffffffffffffffffffffffffffffffffffffffff

\begin{figure*}
\plotfull{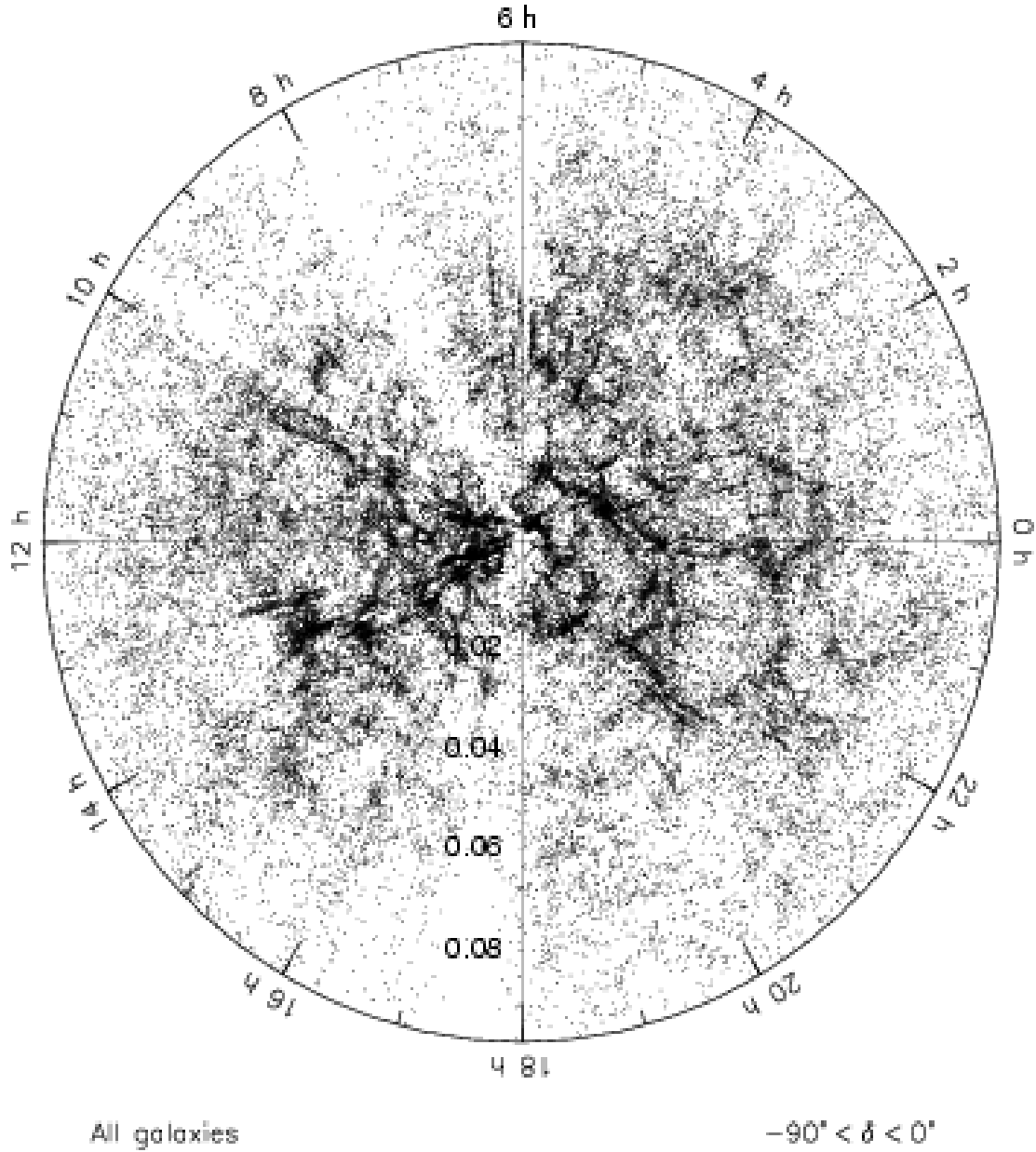}{1.0}
\caption{Spatial redshift distribution combining the 6dF and literature 
redshifts. The redshift slice
  projects through all southerly declinations, $\delta < 0^\circ$.  The
  sparse sampling around 8 and 17~hr is due to non-coverage in the
  Galactic Plane. Variations elsewhere in the sky are due to different
  sky regions having different observational completenesses at the time
  of this First Data Release.}
\label{fig:radplot}
\end{figure*}

\begin{figure*}
\plotfull{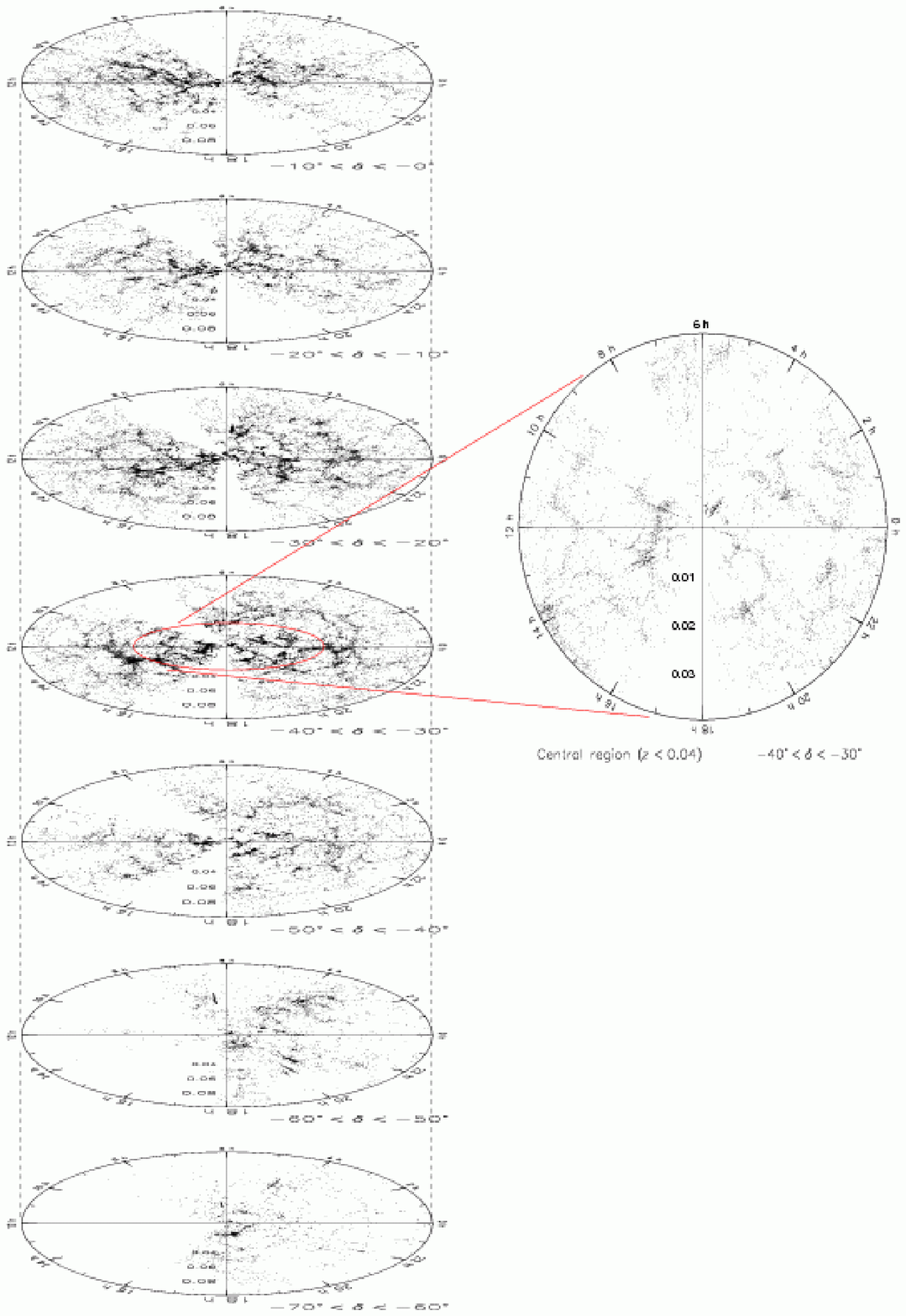}{0.95}
\caption{Spatial redshift distribution divided into discs, each spanning
a $10^\circ$ range in declination. The inset shows an expanded view of
the central region of the $-40^\circ < \delta < -30^\circ$ slice.}
\label{fig:cylplot}
\end{figure*}

Figure~\ref{fig:radplot} shows the radial distribution of galaxies
across the southern sky, projected across the full range of southerly
declinations ($\delta = 0$ to $-90^\circ$). Projecting in this way has
the drawback of taking truly separate 3D space structures and blending
them on the 2D page. Figure~\ref{fig:cylplot} shows the same data
plotted $\Delta \delta = 10^\circ$ declination slices and a magnified
view of the lowest redshift galaxies within $-40^\circ < \delta < -30^\circ$.

Variations in galaxy density apparent in Figs.~\ref{fig:radplot} and
\ref{fig:cylplot} are due the incomplete coverage of observed fields 
and the projection of the Galactic Plane. 
No 6dFGS galaxies lie within galactic latitude
$|\,b\,| \le 10^\circ$. 
% as evident in the two sparsely populated sectors
% at $\sim 8$~hr and $\sim 17$~hr right ascension.  
The 6dFGS is also
clumpier than optically-selected redshift surveys such as 2dFGRS and
SDSS. This is because the near-infrared selection is biased towards
early-type galaxies, which cluster more strongly than spirals.

% ffffffffffffffffffffffffffffffffffffffffffffffffffffffffffffffffffffffffff
% ffffffffffffffffffffffffffffffffffffffffffffffffffffffffffffffffffffffffff

% ffffffffffffffffffffffffffffffffffffffffffffffffffffffffffffffffffffffffff
% ffffffffffffffffffffffffffffffffffffffffffffffffffffffffffffffffffffffffff

The 6dFGS provides the largest sample of near-infrared selected galaxies
to determine the fraction of mass in the present-day universe existing
in the form of stars. To this end, Jones \etal\ (2004) are deriving the
$J$, $H$ and $K_{\rm s}$-band luminosity functions from the first \jonesLF\
redshifts of the 6dF Galaxy Survey, combining data from both before and
after the First Data Release. Using the near-infrared luminosity
functions and stellar population synthesis models, the galaxy
stellar-mass function for the local universe can be estimated. When this
is integrated over the full range of galaxy masses, the total mass of
the present-day universe in stars can be expressed in units of the
critical density.

\subsection{6dFGS Online Database}
\label{sec:database}

Data from the 6dF Galaxy Survey are publicly accessible through an online
database at {\tt http://www-wfau.roe.ac.uk/6dFGS/}, and maintained by
the Wide Field Astronomy Unit of the Institute for Astronomy, University
of Edinburgh. An early data release of around 17\,000 redshifts was made in
December 2002, along with the opening of the web site and tools for
catalogue access. This paper marks the First Data Release of \totalDB\ total
redshifts measured between January 2002 and July 2003. The design of the
database is similar to that used for the 2dF Galaxy Redshift Survey in
that parameterised data are stored in a relational database.  Each {\tt
TARGET} object is also represented by a multi-extension FITS file
which holds thumbnail images of the object and the spectra. The database
is accessed/queried using Structured Query Language (SQL). A combined
6dF-literature redshift catalogue is provided in a separate single
master catalogue.

The 6dFGS database is housed under Microsoft's relational database
software, {\sl SQL Server 2000}. The data are organised in several
tables (Table \ref{tab:database1}). The master target list used to
configure 6dFGS observations is represented by the {\tt TARGET} table.
Spectral observations are stored in the {\tt SPECTRA} table. The input
catalogues that were merged to make up the master target list are also
held in individual tables ({\tt TWOMASS}, {\tt SUPERCOS} etc.). The {\tt
TARGET} table forms the hub of the database. Every table is
interlinked via the parameters targetid and targetname. These parameters
are unique in the master {\tt TARGET} table but are not necessarily
unique in the other tables, ({\em e.g.} {\tt SPECTRA}) as objects can
and have been observed more than once. The {\tt SPECTRA} table holds all
the observational and redshift related data. Parameters are recorded for
both the V and R frames (with a lot of the values being the same for
both frames), and redshift information is derived from the combined VR
frame.  The {\tt TWOMASS} table contains the $K$, $J$ and $H$-selected
samples originating from the 2MASS extended source catalogue. The
$K$-selected sample represents the primary 6dFGS input catalogue.
Table~\ref{tab:database1} lists the programme details for the other
contributing samples.

Initially every FITS file, representing each target ({\tt
targetname.fits}), holds thumbnail images of the target.  As data are
ingested into the database the reduced spectra are stored as additional
FITS image extensions. Table~\ref{tab:database2} summarises the content
within each FITS extension. The first 5 extensions contain the thumbnail
images and each have a built-in World Coordinate System (WCS). The
optical $B$ and $R$ images come from SuperCOSMOS scans of blue (\bj) and
red (\rf) survey plates. The 2MASS $J$, $H$ and $K$ images were
extracted from datacubes supplied by IPAC. Note that although some
objects in {\tt TARGET} do not have 2MASS images, the corresponding
extensions still exist in the FITS file but contain small placeholder
images. The remaining extensions contain the spectra. Each 6dFGS
observation will usually result in a further 3 extensions, the V grating
spectrum, the R spectrum and the combined/spliced VR spectrum.

% ttttttttttttttttttttttttttttttttttttttttttttttttttttttttttttttttttttt
\begin{table*}
\begin{center}
\caption{Tables of data in the 6dFGS Database 
\label{tab:database1}
} 
\vspace{6pt}
\begin{tabular}{llc}
\hline \hline
Table name   &        Description  &   Programme  \\
             &                     &   ID Numbers  \\

\hline

{\tt TARGET}    &   the master target list                  &{\tt progid} \\
{\tt SPECTRA}   &   redshifts and observational data               &  $-$    \\
{\tt TWOMASS}   &   2MASS input catalogue $K$, $H$, and $J$        &  1, 3, 4\\
{\tt SUPERCOS}  &   SuperCOSMOS bright galaxies $b_J$ and $r_F$    &  7, 8   \\
{\tt FSC}       &   sources from the IRAS FAINT Source Catalogue   &  126    \\
{\tt RASS}      &   candidate AGN from the ROSAT All-Sky Survey    &  113    \\
{\tt HIPASS}    &   sources from the HIPASS HI survey              &  119    \\
{\tt DURUKST}   &   extension to Durham/UKST galaxy survey         &  78     \\
{\tt SHAPLEY}   &   galaxies from the Shapley supercluster         &  90     \\
{\tt DENISI}    &   galaxies from DENIS $I < 14.85$                &  6      \\
{\tt DENISJ}    &   galaxies from DENIS $J < 13.85$                &  5      \\
{\tt AGN2MASS}  &   candidate AGN from the 2MASS red AGN survey    &  116    \\
{\tt HES}       &   candidate QSOs from the Hamburg/ESO Survey     &  129    \\
{\tt NVSS}      &   candidate QSOs from NVSS                       &  130    \\
{\tt SUMSS}     &   radio source IDs from SUMSS and NVSS           &  125    \\

\hline \hline
\end{tabular}
\end{center}
\end{table*}
% ttttttttttttttttttttttttttttttttttttttttttttttttttttttttttttttttttttt

The V and R extensions are images with 3 rows. The 1st row is the
observed reduced {\tt SPECTRUM}, the 2nd row is the associated variance
and the 3rd row stores the SKY spectrum as recorded for each data frame.
Wavelength information is provided in the header keywords {\tt CRVAL1},
{\tt CDELT1} and {\tt CRPIX1}, such that
% %%%%%
% \begin{equation}
\begin{eqnarray}
{\rm wavelength\,(\AA)} & = & {\tt CRVAL1} - ({\tt CRPIX1} - 
                                           {\rm pixel\,number}) \nonumber\\
                       &   &  \times \, {\tt CDELT1} .
\end{eqnarray}
\label{crvaleqn}
% \end{equation}
% %%%%%

Additional WCS keywords are also included to ensure the wavelength
information is displayed correctly when using image browsers such as
Starlink's GAIA or SAOimage DS9.

The VR extension also has an additional 4th row that represents the {\tt
WAVELENGTH} axis, which has a continuous dispersion, achieved through
the continuation of the V dispersion into the R half from rescrunching.

% ttttttttttttttttttttttttttttttttttttttttttttttttttttttttttttttttttttt
\begin{table}
\begin{center}
\caption{Contents of each extension in the database FITS files  
\label{tab:database2}
} 
\vspace{6pt}
\begin{tabular}{ll}
\hline \hline
FITS     &  Contents \\
Extension     &           \\

\hline 

1st &   SuperCOSMOS \bj\ image ($1 \times 1$ arcmin) \\
2nd &   SuperCOSMOS \rf\ image ($1 \times 1$ arcmin)  \\
3rd &   2MASS $J$ image (variable size)  \\
4th &   2MASS $H$ image (variable size)  \\
5th &   2MASS $K$ image (variable size)  \\
6th &   V-spectrum extension   \\
7th &   R-spectrum extension   \\
8th &   combined VR-spectrum extension   \\
               &                                     \\
$n$th     &  additional V, R, and VR data \\ 

\hline \hline
\end{tabular}
\end{center}
\end{table}
% ttttttttttttttttttttttttttttttttttttttttttttttttttttttttttttttttttttt

Access to the database is through two different Hypertext Mark-up
Language (HTML) entry forms. Both parse the user input and submit an SQL
request to the database. For users unfamiliar with SQL, the menu driven
form provides guidance in constructing a query. The SQL query box form
allows users more comfortable with SQL access to the full range of SQL
commands and syntax. Both forms allow the user to select different types
of output (HTML, comma separated value (CSV) or a TAR save-set of FITS
files).

There are online examples of different queries using either the menu
or SQL form at {\tt http://www-wfau.roe.ac.uk/6dFGS/examples.html}.
More information about the database is available directly from the
6dFGS database website.

% ssssssssssssssssssssssssssssssssssssssssssssssssssssssssssssssssssssss

\section{CONCLUSIONS}
\label{sec:conclusions}

The 6dF Galaxy Redshift Survey (6dFGS) is designed to measure redshifts
for approximately \aimsixdf\ galaxies and the peculiar velocities of
15\,000. The survey uses the 6dF multi-fibre spectrograph on the United
Kingdom Schmidt Telescope, which is capable of observing up to 150
objects simultaneously over a $5.7^\circ$-diameter field of view.  The
2MASS Extended Source Catalog (Jarrett \etal\ 2000) is the primary
source from which targets have been selected. The primary sample has
been selected with $K_{\rm tot} \leq 12.75$, where $K_{\rm tot}$ denotes
the total $K$-band magnitude as derived from the isophotal 2MASS $K$
photometry. Additional galaxies have been selected to complete the
target list down to $(H, J, r_F, b_J) = (13.05, 13.75, 15.6, 16.75)$.
Thirteen miscellaneous surveys complete the total target list.

The survey covers the entire southern sky (declination $\delta <
0^\circ$), save for the regions within $|\,b\,| \leq 10^\circ$ of the
Galactic Plane. This area is has been tiled with around 1500 fields that
effectively cover the southern sky twice over. An adaptive tiling
algorithm has been used to provide a uniform sampling rate of 94\%. In
total the survey covers some 17\,046\,deg$^2$ and has a median depth of
$\bar{z}$=0.05. There are three stages to the observations, which
initially target the declination strip $-42^\circ<\delta<-23^\circ$,
followed by the equatorial region $-23^\circ<\delta<0^\circ$, and
conclude around the pole, ($\delta<-42^\circ$).

Spectra are obtained through separate V and R gratings and later
spliced to produce combined spectra spanning 4000 -- 8400\,\AA. The
spectra have 5 -- 6\,\AA\ FWHM resolution in V and 9 -- 12\,\AA\ 
resolution in R.  Software is used to estimate redshifts from both
cross-correlation with template absorption-line spectra, and linear fits
to the positions of strong emission lines. Each of these automatic
redshift estimates is checked visually and assigned a quality $Q$ on a
scale of 1 to 5, where $Q \ge 3$ covers the range of reliable redshift
measurements. The median signal-to-noise ratio is 9.4~\perpix\ for
redshifts with quality $Q=4$, and 5.3~\perpix\ for $Q=3$ redshifts.
  
The data in this paper constitute the First Data Release of \totalDB\
observed spectra and the \uniqueDB\ unique extragalactic redshifts from 
this set. The rates of contamination by Galactic and failed spectra are 4\% 
and 11\% respectively. Data from the 6dF Galaxy Survey are publicly
available through an online database at {\tt
  http://www-wfau.roe.ac.uk/6dFGS/}, searchable through either SQL query
commands or a online WWW form.  The main survey web site can be found at
{\tt http://www.mso.anu.edu.au/6dFGS}.

% ssssssssssssssssssssssssssssssssssssssssssssssssssssssssssssssssssssss

\section*{Acknowledgements}

We acknowledge the efforts of the staff of the Anglo-Australian
Observatory, who have undertaken the observations and developed the 6dF
instrument. We are grateful to P.\ Lah for his help in creating
Fig.~\ref{fig:radplot}. D.\ H.\ Jones is supported as a Research
Associate by Australian Research Council Discovery--Projects Grant
(DP-0208876), administered by the Australian National University.

T.\ Jarrett and J.\ Huchra acknowledge the support of NASA. They are
grateful to the other members of 2MASS extragalactic team, M.\ 
Skrutskie, R.\ Cutri, T.\ Chester and S.\ Schneider for help in
producing the major input catalog for the 6dFGRS. They also thank NASA,
the NSF, the USAF and USN and the State of Massachussetts for the
support of the 2MASS project and NASA for the support of the 6dF
observational facility.

The DENIS project has been partly funded by the SCIENCE and the HCM
plans of the European Commission under grants CT920791 and CT940627. It
is supported by INSU, MEN and CNRS in France, by the State of
Baden-Warttemberg in Germany, by DGICYT in Spain, by CNR in Italy, by
FFwFBWF in Austria, by FAPESP in Brazil, by OTKA grants F-4239 and
F-013990 in Hungary, and by the ESO C\&EE grant A-04-046.

% bbbbbbbbbbbbbbbbbbbbbbbbbbbbbbbbbbbbbbbbbbbbbbbbbbbbbbbbbbbbbbbbbbbbbb


\begin{thebibliography}{}
\bibitem{}
Blanton, M.R. \etal, 2001, AJ, 121, 235
\bibitem
Blanton, M.R. \etal, 2003, AJ, 125, 2276
\bibitem{}
Branchini, E. \etal, 1999, MNRAS, 308, 18
\bibitem{}
Burkey, D., 2004, PhD dissertation, in prep.
\bibitem{}
Burkey, D. \& Taylor, A., 2004, MNRAS, submitted
\bibitem{}
Campbell, L.A. \etal, 2004, MNRAS, in press 
\bibitem{}
Cole S. \etal, (2dFGRS team), 2001, MNRAS, 326, 255
\bibitem{}
Colless, M.M. \etal, 2001a, MNRAS, 321, 277
\bibitem{}
Colless, M.M. \etal, (2dFGRS team), 2001b, MNRAS, 328, 1039
\bibitem{}
Cross, N. \etal, (2dFGRS team), 2001, MNRAS, 324, 825
\bibitem{}
da Costa, L.N. \etal, 2000, ApJ, 537, L81
\bibitem{}
De Propris, R. \etal, (2dFGRS team), 2002, MNRAS, 329, 87
\bibitem{}
Djorgovski, S. \& Davis, M., 1987, ApJ, 313, 59 
\bibitem{}
Dressler, A. \etal, 1987, ApJ, 313, 42
\bibitem{}
Efstathiou, G. \etal, (2dFGRS team), 2002, MNRAS, 330, 29
\bibitem{}
Folkes, S. \etal, (2dFGRS team), 1999, MNRAS, 308, 459
\bibitem{}
Giovanelli, R. \etal, 1998, AJ, 116, 2632
\bibitem{}
Goto T. \etal, 2003, PASJ, 55, 739
\bibitem{}
Hambly, N.C. \etal, 2001, MNRAS, 326, 1279 
\bibitem{}
Hawkins, E. \etal, (2dFGRS team), 2003, MNRAS, 346, 78 
\bibitem{}
Hoeg, E. \etal, 2000, A\&A, 355, L27
\bibitem{}
Huchra, J. \etal, ApJS, 121, 287
\bibitem{}
Hudson, M.J. \etal, 1999, ApJ, 512, L79 
\bibitem{}
Jarrett, T.-H. \etal, 2000, AJ 120, 298
\bibitem{}
Jones, D.H. \etal, 2004, in prep. 
\bibitem{}
Lahav, O. \etal, (2dFGRS team), 2002, MNRAS, 333, 961
\bibitem{}
Lauer, T.R. \& Postman, M., 1994, ApJ, 425, 418
\bibitem{}
Lewis, I.J. \etal, 2002, MNRAS, 333, 279
\bibitem{}
Lynden-Bell, D. \etal, 1988, ApJ, 326, 19
\bibitem{}
Madgwick, D.S. \etal, (2dFGRS team), 2002, MNRAS, 333, 133
\bibitem{}
Metropolis, N. \etal, 1953, J.\ Chem.\ Phys., 21
\bibitem{}
Norberg, P. \etal, 2002, MNRAS, 336, 907
\bibitem{}
Parker, Q.A. \etal, 1998, in
{\em Fiber Optics in Astronomy III}, ASP Conf Series 152, p80
\bibitem{}
Parker, Q.A \& Watson, F.G., 1995, in {\em Fiber Optics in Astronomical
Applications}, Proc SPIE v2476, ed. S. Barden, p34
\bibitem{}
Peacock, J.A. \etal, (2dFGRS team), 2001, Nature, 410, 169
\bibitem{}
Percival W.J. \etal, (2dFGRS team), 2001, MNRAS, 327, 1297
\bibitem{}
Saunders, W. \etal, 2000, MNRAS, 317, 55
\bibitem{}
Saunders W. \etal, 2001, {\em AAO Newsletter}, 97, 14
\bibitem{}
Scaramella, R. \etal, Nature, 338, 562
\bibitem{}
Szalay, A. \etal, 2003, ApJ, 591, 1
\bibitem{}
Verde, L. \etal, (2dFGRS team), 2002, MNRAS, 335, 432
\bibitem{}
Watson, F.G \etal, 2000, in {\em Optical and
IR Telescope Instrumentation and Detectors}, Proc SPIE vol 4008,
eds. M. Iye, A.F. Moorwood, p123
\bibitem{}
Wegner, G.A. \etal, 1999, MNRAS, 305, 259
\bibitem{}
York, D.G. \etal, 2001, AJ 120, 1579
\bibitem{}
Zehavi, I. \etal, (SDSS team), 2002, ApJ, 571, 172

\end{thebibliography}
\end{document}